\documentclass[twocolumn,reprint,superscriptaddress,amsmath,amssymb,aps,nofootinbib]{revtex4-1}

\usepackage{graphicx}
\usepackage{dcolumn}
\usepackage{bm}
\usepackage{hyperref}
\usepackage[utf8]{inputenc}
\usepackage[dvipsnames]{xcolor}
\usepackage[normalem]{ulem}
\usepackage{xspace}
\usepackage{fontawesome} 
\usepackage{multirow}

\newcommand{\pbh}{{\mathrm{PBH}}}
\newcommand{\OmegaPBH}{\Omega_\mathrm{PBH}}
\newcommand{\dm}{{\chi}}
\newcommand{\cdm}{{\mathrm{DM}}}
\newcommand{\Fermi}{\textit{Fermi}\xspace}
\newcommand{\us}[1]{{\mathrm{~#1}}} 
\renewcommand\u[1]{{\mathrm{#1}}} 
\newcommand{\ee}[1]{\times10^{{#1}}} 
\newcommand{\ev}[1]{{\langle {#1} \rangle}}
\newcommand{\sv}{(\sigma v_\text{rel})_0}

\newcommand{\VT}{\ev{VT}}
\newcommand{\Rbar}{\overline{\mathcal{R}}}

\newcommand{\eV}{\text{e\kern-0.2ex V}\xspace}

\newcommand{\TeV}{\text{T\kern-0.1ex \eV}\xspace}

\newcommand\myshade{80}
\colorlet{mylinkcolor}{ForestGreen}
\colorlet{mycitecolor}{Red}
\colorlet{myurlcolor}{violet}

\hypersetup{
  linkcolor  = mylinkcolor!\myshade!black,
  citecolor  = mycitecolor!\myshade!black,
  urlcolor   = myurlcolor!\myshade!black,
  colorlinks = true
}

\newcommand{\GRAPPA}{%
Gravitation Astroparticle Physics Amsterdam (GRAPPA),\\
Institute for Theoretical Physics Amsterdam
and Delta Institute for Theoretical Physics,\\
University of Amsterdam, Science Park 904, 1098 XH Amsterdam, The Netherlands}

\begin{document}

\title{Primordial Black Holes as Silver Bullets for New Physics at the Weak Scale}

\author{Gianfranco Bertone}%
\email{g.bertone@uva.nl}
\affiliation{\GRAPPA}%
\author{Adam M. Coogan}%
\email{a.m.coogan@uva.nl}
\affiliation{\GRAPPA}%
\author{Daniele Gaggero}%
\email{daniele.gaggero@uam.es}
\affiliation{Instituto de F\'isica Te\'orica UAM/CSIC,
Calle Nicol\'as Cabrera 13-15, Cantoblanco E-28049 Madrid, Spain}%
\author{Bradley J. Kavanagh}%
\email{b.j.kavanagh@uva.nl}
\affiliation{\GRAPPA}
\author{Christoph Weniger}%
\email{c.weniger@uva.nl}
\affiliation{\GRAPPA}%

\date{\today}

\begin{abstract}
Observational constraints on gamma rays produced by the annihilation of weakly interacting massive particles around primordial black holes (PBHs) imply that these two classes of Dark Matter candidates cannot coexist. We show here that the successful detection of one or more PBHs by radio searches (with the Square Kilometer Array) and gravitational waves searches (with LIGO/Virgo and the upcoming Einstein Telescope) would set extraordinarily stringent constraints on virtually all weak-scale extensions of the Standard Model with stable relics, including those predicting a WIMP abundance much smaller than that of Dark Matter. Upcoming PBHs searches have in particular the potential to rule out {\it almost the entire} parameter space of popular theories such as the minimal supersymmetric standard model and scalar singlet Dark Matter.
\href{https://github.com/adam-coogan/pbhs_vs_wimps}{\faGithub} \href{https://doi.org/10.5281/zenodo.2662141}{\faTags}
\end{abstract}

\maketitle

\paragraph*{Introduction.}
The formation and growth of black holes (BHs) inevitably modifies the Dark Matter (DM) distribution around them. If DM is in the form of weakly interacting massive particles (WIMPs) which self-annihilate, the increase in DM density can significantly boost the annihilation rate. This process has been discussed in the context of supermassive BHs at the center of galaxies \cite{Gondolo:1999ef,Gondolo:2000pn,Ullio:2001fb,Bertone:2001jv,Merritt:2002vj,Bertone:2005hw,Merritt:2006mt} and intermediate-mass BHs \cite{Bertone:2005xz,Zhao:2005zr,Bringmann:2009ip}.
The argument has been more recently extended to the case of {\it primordial} black holes (PBHs), which can form before Big Bang nucleosynthesis \cite{Hawking:1971ei,Carr:1974nx} and could constitute a significant, yet subdominant, fraction of DM. 
In this case, the WIMP annihilation rate around PBHs would lead to a gamma-ray background exceeding the one observed by the \Fermi Large Area Telescope (LAT), leading to stringent constraints on the relative PBH abundance $f_\pbh = \OmegaPBH/\Omega_\cdm$~\cite{Lacki:2010zf,Eroshenko:2016yve,Boucenna:2017ghj,Adamek:2019gns}.

In this \emph{letter}, we explore the compatibility of PBHs and WIMP DM from the opposite viewpoint.  We focus on the prospects for discovering PBHs with upcoming radio and gravitational wave (GW) searches, and on the implications such a discovery would have on even a small relic density of WIMPs in the Universe.  Specifically, we consider three discovery scenarios: {\it i)} The detection of GWs produced by the merger of BHs with mass $M \lesssim 1 \,M_\odot$ with LIGO/Virgo; {\it ii)} The detection of GWs produced by the merger of $\mathcal{O}(10\, M_\odot)$ BHs at redshift $z > 40$ with the Einstein Telescope; {\it iii)} The detection of the radio emission produced by the accretion of gas onto 1--1000 M$_\odot$ BHs with the planned Square Kilometer Array (SKA). The scenarios we consider are summarized in Table~\ref{tab:Detections}.

\begin{figure}[tbh!]
\centering
   \includegraphics[width=\linewidth,]{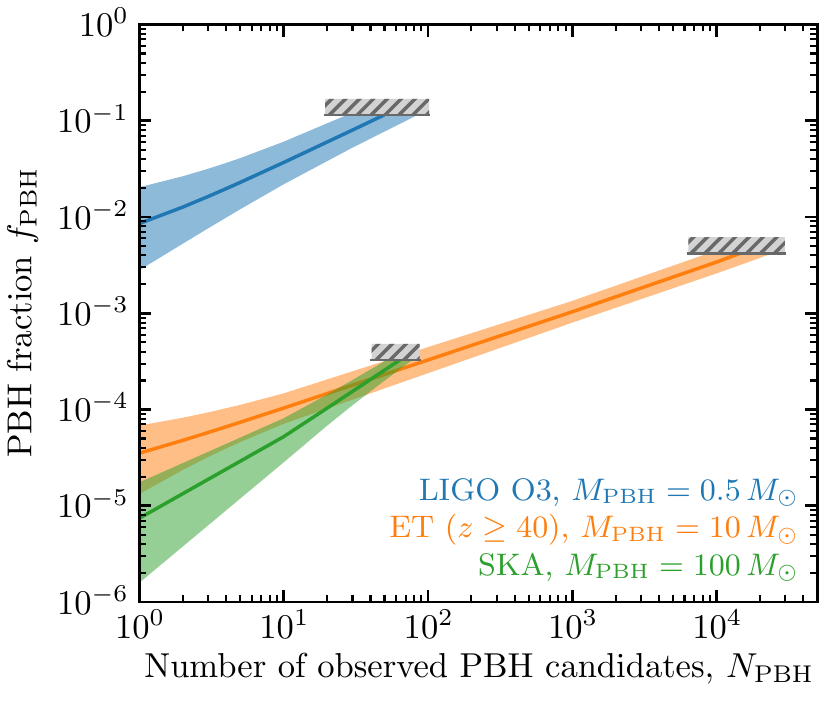}
   \caption{{\bf Constraints on the PBH fraction from future observations of PBHs.} Median value (solid line) and symmetric 90\% credible intervals \cite{Aasi:2013jjl} (shaded region) of $f_\pbh = \OmegaPBH/\Omega_\cdm$, assuming the observation of $N$ PBH candidates. In blue, we assume the observation of BH mergers with component masses of $0.5 \, M_\odot$ during LIGO O3. In orange, we assume BH mergers with component masses of $10\, M_\odot$ are observed at redshift $z \geq 40$ during 1 year of operation of Einstein Telescope. In green, we assume the observation of 100 $M_\odot$ PBHs in the Milky Way in radio and X-ray searches. The grey hatched boundaries show the current 95\% upper limit on $f_\pbh$ for each PBH mass. \href{https://github.com/adam-coogan/pbhs_vs_wimps/blob/master/plot_pbh_fraction.py}{\faFileCodeO}
   }
   \label{fig:Constraints_f}
\end{figure}

We estimate the abundance of PBHs in each scenario, given a number of detections (Fig.~\ref{fig:Constraints_f}) and, from that, we calculate the gamma-ray luminosity of WIMP overdensities around PBHs in the Universe. By comparing this with the observed diffuse extragalactic gamma-ray flux and with unidentified gamma-ray point sources in the 3FGL \Fermi-LAT catalogue, we show that a positive detection of even a small number of PBHs in any of the above scenarios would set extraordinarily stringent constraints on weak-scale extensions of the Standard Model with stable relics, including those predicting a WIMP abundance much smaller than that of the total Dark Matter.

Our code is available on GitHub \href{https://github.com/adam-coogan/pbhs_vs_wimps}{\faGithub}, and a link below each figure (\faFileCodeO) provides the python code with which it was generated. An archived version is available on Zenodo \href{https://doi.org/10.5281/zenodo.2662141}{\faTags}.

\paragraph*{Discovering PBHs with Gravitational Waves.}
Upcoming GW observations have the potential to unequivocally detect PBHs.  We consider two main detection scenarios based on this channel. The first case is the observation of nearby mergers of sub-solar mass BHs. The lightest known BH is in the X-ray binary XTE J1650-500, with a mass of $5.1\,M_\odot$ \cite{Slany:2008pm} or perhaps larger \cite{Shaposhnikov:2009dc}. In standard theories of stellar evolution, compact objects do not form below the Chandrasekhar mass of $\sim 1.4\,M_\odot$ \cite{1931ApJ....74...81C,1931MNRAS..91..456C,1935MNRAS..95..207C}.  As a result, the detection of sub-solar mass BHs would strongly suggest a primordial origin (though see Refs.~\cite{Shandera:2018xkn,Kouvaris:2018wnh} for alternative exotic formation mechanisms). 
The second case is the observation of BH-BH mergers at high redshift with proposed detectors such as Einstein Telescope (ET) \cite{Hild:2010id} and Cosmic Explorer (CE) \cite{Sathyaprakash:2012jk}. At low redshift, mergers of astrophysical BHs are commonplace, but at earlier times BHs have typically not had sufficient time to collapse and form binaries \cite{Mapelli:2019bnp}. Above around $z \sim 40$, the merger rate of astrophysical BHs should be negligible \cite{Koushiappas:2017kqm}. High-redshift mergers therefore point towards a primordial BH population which formed much earlier \cite{Chen:2019irf}.

\paragraph*{PBH Merger Rates: Setting the stage.}
To calculate the PBH merger rate as function of the PBH abundance, we follow Refs.~\cite{Ali-Haimoud:2017rtz, Kavanagh:2018ggo} (though see also Refs.~\cite{Hayasaki:2009ug,Raidal:2017mfl,Chen:2018czv,Ballesteros:2018swv,Belotsky:2018wph,Bringmann:2018mxj,Raidal:2018bbj}). In this scenario, pairs of PBHs decouple from the Hubble flow and form binaries before matter-radiation equality \cite{Nakamura:1997sm,Ioka:1998nz,Sasaki:2016jop}.
For a given value of $(f_\pbh, M_\pbh)$, we calculate the comoving PBH merger rate (in the source frame) as a function of redshift as:
\begin{equation}
\label{eq:Rmerge}
\mathcal{R}(z) = \frac{1}{2}n_\pbh P(t_\mathrm{merge} = t[z])\,.
\end{equation}
Here, $n_\pbh =  f_\pbh \Omega_\cdm \rho_{c, 0}/M_\pbh$ is the comoving number density of PBHs in terms of the critical density today $\rho_{c, 0}$. The merger time distribution is $P(t_\mathrm{merge})$, which we evaluate at $t_\mathrm{merge} = t[z]$, the age of the Universe at redshift $z$. 
The merger time depends on the semi-major axis $a$ and eccentricity $e$ of the binary orbit \cite{Peters:1964zz}. The distribution $P(t_\mathrm{merge})$ is thus obtained from the distribution of $P(a, e |M_\pbh)$, calculated assuming a uniform initial distribution of PBHs and accounting for torques due to all other PBHs and smooth density perturbations \cite{Ali-Haimoud:2017rtz}.  We also account for the impact of local DM halos on the binary properties \cite{Kavanagh:2018ggo}, for which we assume a DM density profile with slope $\gamma = 9/4$ \cite{Adamek:2019gns}, leading to a roughly 30\% enhancement in the merger rate. 


{\renewcommand{\arraystretch}{1.2}
\begin{table}[tbh!]
    \centering
    \begin{tabular}{c|c|c|c}
        \toprule
         & Sub-solar (O3) & High redshift (ET) & SKA\\
        \colrule
        $M_\pbh$ [$M_\odot$]  & 0.5 & 10 & 100\\
        $N_\u{min}$  & 1 & 1 & 10\\
        $N_\u{max}$ & 80 & 24000 & 80 \\
        \botrule
    \end{tabular}
    \caption{\textbf{PBH Detection Scenarios.} For each of the three detection scenarios we consider, we show the benchmark PBH mass $M_\odot$ which we assume, as well as $N_\mathrm{min}$ and $N_\mathrm{max}$, the smallest and largest number of detections we consider. Note that $N_\mathrm{max}$ is the largest number of detections which would be compatible with current PBH constraints.}
    \label{tab:Detections}
\end{table}}

To obtain projected constraints on $\mathcal{R}$ (or equivalently $f_\pbh$) from future GW observation campaigns, we follow LIGO/Virgo \cite{Abbott:2016nhf,Abbott:2016drs} and perform a Bayesian analysis\footnote{We will assume that the threshold for detection of mergers is high enough that there is no contamination from terrestrial `events' (i.e. $p_\mathrm{astro} \simeq 1$ for all observations).}. We assume that the number of observed mergers is Poisson distributed, $P(N|\Lambda) = \mathrm{Poiss}(\Lambda)$, with mean: 
\begin{align}
\Lambda = \int_{z_1}^{z_2} \mathcal{R}( z )\frac{\mathrm{d}\VT}{\mathrm{d}z} \, \mathrm { d } z  \equiv \Rbar \langle VT \rangle\, ,
\label{eq:lambda}
\end{align}
where $\mathrm{d}\VT / \mathrm{d}z = T \frac { 1 } { 1 + z } \frac { \mathrm { d } V _ { c } } { \mathrm { d } z } f ( z )$ is the differential time-volume sensitivity for an observation time $T$~\cite{Abbott:2016drs,Abbott:2016nhf}.
We calculate the comoving volume as a function of redshift $\mathrm{d}V_c/\mathrm{d}z$ \cite{Hogg:1999ad} assuming the Planck 2018 best fit cosmological parameters~\cite{Aghanim:2018eyx}.
The selection function $f(z)$ encodes the detection probability for a given source at redshift $z$.

As in Refs.~\cite{Abbott:2016nhf,Abbott:2016drs}, we assume a Jeffreys' prior\footnote{The Jeffreys prior is an uninformative prior which is invariant under reparametrization of the parameter of interest \cite{Jeffreys}.} on $\Lambda$ (though see Appendix~\ref{app:prior_dependence} for a more detailed discussion of the prior dependence of our results). Using Bayes' Theorem \cite{Bayes:1764vd}, the resulting posterior on $\Lambda$, given $N$ observations, is:
\begin{align}
    P(\Lambda|N)\propto P(\Lambda)P(N|\Lambda) \propto\frac{1}{\Lambda^{1/2}}\Lambda^N \exp\left(-\Lambda\right)\,.
\end{align}
The corresponding posterior on the redshift-averaged merger rate $\Rbar = \Lambda/\VT$ is then:
\begin{align}
\begin{split}
    P(\Rbar|N, \VT) 
    &\propto \VT \left(\Rbar \VT\right)^{N-1/2}\exp\left(-\Rbar \VT\right)\,.
\end{split}
\end{align}
We also incorporate a log-normal prior on the total time-volume sensitivity $\VT$, with a 30\% relative uncertainty in order to account for possible calibration uncertainties \cite{Abbott:2016drs}. We marginalize over $\VT$ to obtain the marginal posterior for $\Rbar$, which can be transformed into a posterior on $f_\pbh$:
\begin{equation}
    P(f_\pbh|N) = P(\Rbar|N)\frac{\partial{\Rbar}}{\partial f_\pbh}\,.
\end{equation}

We now apply this formalism to the specific cases of sub-solar mass PBHs observed with LIGO and high-redshift mergers observed with ET. 

\paragraph*{Sub-solar mass mergers.}
Let us assume that during a one-year campaign, LIGO's third observing run (O3) observes $N$ events which are consistent with merging sub-solar mass BHs. Let us also assume that the PBH mass function is monochromatic and that the component masses of the merging BHs are recovered exactly. In this case, we choose as a benchmark a PBH mass of $M_\pbh = 0.5\,M_\odot$.

LIGO O3 is sensitive to such mergers only in a small range of redshifts, $z \lesssim 0.02$. We therefore assume that the PBH merger rate is constant over this range, replacing it with the redshift-averaged value $\Rbar$. We estimate the sensitive time-volume using the approximation of Ref.~\cite{Magee:2018opb}:
\begin{equation}
    \VT  = \frac { 4 \pi} { 3 } D _ { \mathrm{avg}} ^ { 3 } T.
\end{equation}
Here, $T$ is the analyzable live-time of the detectors and $D _ { \mathrm{avg} } $ is the average detector range. This is obtained from $D_\mathrm{max}$, the horizon distance of the detector, by averaging the detector response over both orientation and location of the binary, which gives $D _ \mathrm{avg } \approx D _ { \max } / 2.26$. We assume a duration of one year for O3 with a 70\% duty cycle \cite{Aasi:2013wya}. We also assume a further reduction of 5\% due to data quality cuts \cite{TheLIGOScientific:2017lwt}. The LIGO O2 horizon distance for observing 0.5 $M_\odot$ mergers (with a signal-to-noise of at least 8 in each of the two detectors) is $\sim 60 \,\mathrm{Mpc}$ \cite{Abbott:2018oah}. 
From Ref.~\cite{Aasi:2013wya}, we expect an improvement in the horizon distance of about 50\% from O2 to O3. We therefore assume a horizon distance $D_\mathrm{max}  = 90\,\mathrm{Mpc}$ ($z\approx 0.0192$), giving a sensitive time-volume of $\langle V T \rangle \approx 1.8 \times 10^{-4} \,\mathrm{Gpc}^3 \,\mathrm{yr}$.

The resulting posteriors on $f_\pbh$ are shown in blue in Fig.~\ref{fig:Constraints_f}. 
The grey shaded region shows the current constraint on $0.5\,M_\odot$ PBHs, coming from the observed merger rate during LIGO O2 \cite{Abbott:2018oah}. We note that the median value of $f_\pbh$ scales as $\sqrt{N}$. This is because the merger rate scales predominantly as $f_\pbh^2$, reflecting the number of PBH binaries which are formed in the early Universe. 

\paragraph*{High-redshift mergers.}
In the case of high-redshift mergers, we consider the sensitivity of the proposed Einstein Telescope (ET) \cite{Hild:2010id} with an observing time of one year. We assume that $N$ BH-BH merger events are observed at redshifts greater than $z > 40$. We again assume a mono-chromatic mass function, now at $M_\pbh = 10\,M_\odot$, close to the maximum sensitivity of ET.
We estimate the mean number of observed merger events by integrating Eq.~\eqref{eq:lambda} over $z > 40$, incorporating the full redshift-dependence of the merger rate $\mathcal{R}(z)$. The differential time-volume sensitivity $\mathrm{d}\VT / \mathrm{d}z$ is calculated as in Eq.~\eqref{eq:lambda} with the selection function $f(z)$ estimated from Ref.~\cite{Hall_LIGOdoc,Hall_code}: this corresponds to $\sim 5\%$ of mergers detectable at $z = 40$, decreasing to zero close to $z = 100$.

The resulting posteriors on $f_\pbh$ are shown in orange in Fig.~\ref{fig:Constraints_f}, where we again cut off the posterior at the current upper limit (grey) from Ref.~\cite{Kavanagh:2018ggo}.

\paragraph*{Discovering Galactic PBHs with SKA.}
\label{sec:ska}

Isolated black holes can in principle be detected by observing the  broad-band spectrum of radiation --- spanning from radio waves all the way up to the X-ray band --- produced by the interstellar gas accreting onto them~\cite{Fender:2013ei}. This strategy can in particular be applied to the search for PBHs in the inner Galaxy and results in a bound on their abundance, as shown in~\cite{Gaggero:2016dpq,Ivanov:2019wkq}. 
A more refined analysis based on state-of-the-art numerical simulation of gas accretion onto BHs recently allowed an even stronger bound to be placed on the abundance of a population of PBHs in the $1 - 100$ M$_\odot$ mass window~\cite{2018arXiv181207967M}.

During the next decade, the Square Kilometre Array (SKA) observatory will allow us to perform radio observations with unprecedented sensitivity. In particular, it will become possible to search for a subdominant population of PBHs that amount to a small fraction (even less than percent) of the Dark Matter in the Universe~\cite{Bull:2018lat}. 

A complex data analysis will be needed to identify a possible population of black holes of primordial origin among the plethora of radio sources that SKA will detect. This will be based on a comparison with other wavelengths and on further considerations on the spatial distribution, mass functions and other quantities that may allow us to disentangle a population of conventional astrophysical BHs from PBHs ~\cite{Gaggero:2016dpq,2018arXiv181207967M,Bull:2018lat,Ivanov:2019wkq}.  
Here we assume that, as a result of such data analysis, a certain number of radio sources observed by SKA is clearly associated to a population of PBHs, and we apply a procedure similar to the one described above to infer the cosmic abundance of PBHs from the number of such objects identified in SKA data. We focus in particular on the mid-frequency instrument (SKA1-Mid), which will observe the sky in the $350 \,{\rm MHz}$--$14 \,{\rm GHz}$ domain. We consider $100$ M$_\odot$ PBHs because these massive objects would be more easily detected, given the $\propto M^2$ scaling of the accretion rate, and more easily distinguished from a population of astrophysical BHs. The region of interest is  the Galactic Ridge ($-0.9^{\circ} < l < 0.7^{\circ};  -0.3^{\circ} < b < 0.3^{\circ}$), where a large concentration of molecular gas is present. 

As a first step, we determine the distribution $P(N_\u{SKA}|f_\pbh)$ of the number of PBHs detectable by SKA1-MID (1.4 GHz), adopting the Monte Carlo simulations described in \cite{Gaggero:2016dpq,2018arXiv181207967M}, and assuming gain G = 15 K/Jy, receiver temperature $T_{\rm rx}$ = 25 K, sky temperature towards the GC $T_{\rm sky}$ = 70 K, bandwidth $\Delta\nu$ = 770 MHz, and $\simeq 1000$ h exposure, which corresponds to $85$ nJy sensitivity.

Then, we apply the Bayesian inversion discussed above to compute the posterior distribution function $P(f_\pbh| N_\u{SKA})$
where we again assume a Jeffreys' prior on the mean number of expected observations.
We show in green in Fig.~\ref{fig:Constraints_f} the resulting posteriors on $f_\pbh$. The current upper bound for $100\,M_\odot$ PBHs comes from the radio and X-ray searches mentioned above \cite{2018arXiv181207967M}.

\paragraph*{Constraints on DM self-annihilation from gamma rays.}
\label{sec:gamma}

Primordial black holes seed the growth of very steep Dark Matter halos that form via approximately radial infall. The theory of secondary infall~\cite{Bertschinger:1985pd} and simple 1D simulations~\cite{Mack:2006gz} predict that these \emph{ultra-compact minihalos} (UCMHs) have $\rho(r) \propto r^{-9/4}$ density profiles, which has been confirmed by recent 3D simulations~\cite{Adamek:2019gns}. 
Since $f_\pbh$ is at or well below the percent-level in all but one of our detection scenarios, we can assume that UCMHs form in isolation, so we neglect the effects of PBH-PBH interactions on the UCMH profile.

Due to the steepness of the profile the WIMP density reaches a maximum value at the ``annihilation plateau'', where the WIMP annihilation rate becomes equal to the Hubble rate. Due to the large resulting gamma-ray luminosities, UCMHs in the Milky Way would appear as bright point sources with no counterparts in other wavelengths. Previous analyses searching the 3FGL for WIMP subhalos~\cite{Bertoni:2015mla,Hooper:2016cld,Schoonenberg:2016aml} have identified 19 bright, high-latitude, non-variable unassociated point sources that are spectrally compatible with annihilating WIMPs. As described in detail in Appendix~\ref{app:gamma_ray_constraints}, we perform a Monte Carlo simulation to assess the observability of UCMHs by Fermi. We then use this to determine the 95\% confidence level (CL) upper bound on  the  WIMP annihilation  cross-section  in the zero-velocity  limit $\sv$. This upper limit  depends on the PBHs' spatial distribution which we assume tracks the Milky Way DM distribution. We fix $f_\pbh$ to the 5th percentile of the posterior $P(f_\pbh | N)$,  derived in the previous sections for the detection of $N$ PBH candidates.  We conservatively assume that all 19 compatible unassociated point sources are UCMHs and set the upper limit on $\sv$ by comparing with the expected number of UCMHs passing cuts on their integrated gamma-ray flux and galactic latitude  (given $M_\pbh$, $m_\dm$ and $N$).

Annihilation in UCMHs outside the Milky Way over all redshifts contributes to the diffuse, isotropic extragalactic background (EGB)~\cite{Taylor:2002zd,Ullio:2002pj,Ando:2005hr}, which has been measured by Fermi~\cite{Ackermann:2014usa}. This provides an additional very robust constraint on the DM self-annihilation cross section since it requires no assumptions about the PBH spatial distribution. To set a conservative bound we do not assume a particular background model. Instead, we compute the expected gamma-ray flux from UCMHs in each of Fermi's energy bins, and calculate the likelihood of such an excess above the observed flux using the statistical and systematic uncertainties. As for the point source constraints, we fix $f_\pbh$ to the 5th percentile for a given detection scenario.

An important difference with regard to standard indirect detection analyses is the scaling of signals with the fractional WIMP abundance $f_\dm = \Omega_\dm / \Omega_\cdm$ for under-abundant thermal relics. Typically, the DM annihilation rate depends on the combination $f_\dm{}^2 \sv$ since it factors into terms dependent on the integrated DM density profile squared ($J$-factor) and the self-annihilation cross section. In the PBH scenario, the DM density profile itself depends on $\sv$ since this sets the radius of the annihilation plateau. As a result, the DM annihilation rate (and thus the extragalactic diffuse flux from PBHs and expected number of unassociated point sources) depends on the combination $f_\dm{}^4 \sv$; this is derived in Appendix~\ref{app:gamma_ray_constraints}. 

\paragraph*{Results and discussion.}
\label{sec:res}

\begin{figure}
    \centering
    \includegraphics[width=\linewidth]{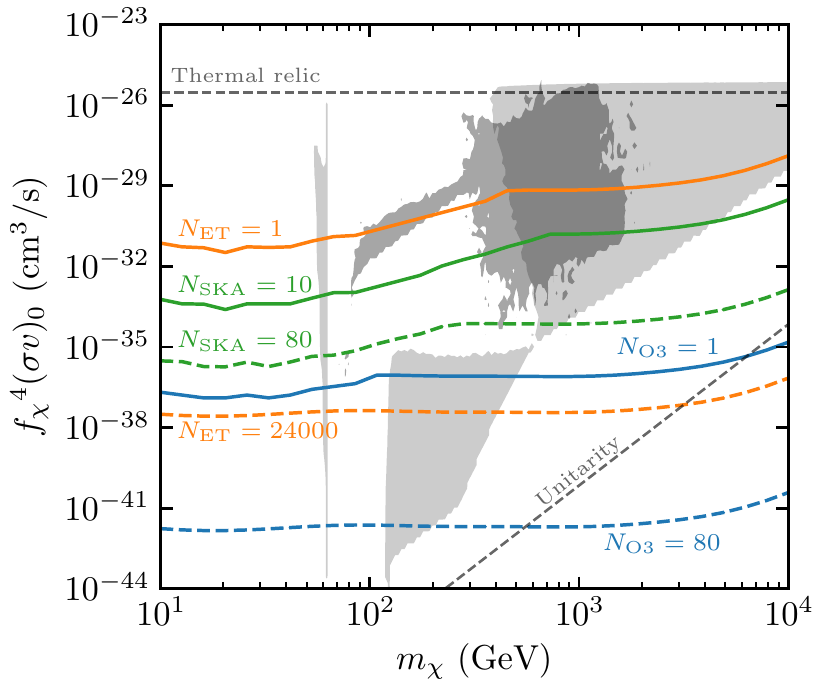}
    \caption{\textbf{Constraints on DM self-annihilation cross section.} The solid lines correspond to the 95\% CL upper limits obtained assuming a small number of PBH detections with LIGO/Virgo O3 (blue), Einstein Telescope (ET, orange) and SKA (green). The lower dashed lines correspond to constraints which would be obtained if the number of PBH observations are as large as allowed by current limits. The dark grey region is the envelope of 95\% CL profile likelihood contours for several supersymmetric models, while the light grey region is for singlet scalar scenarios. The horizontal dotted black line indicates the standard thermal relic cross section $3\ee{-26}\us{cm}^3/\u{s}$. The angled dotted black line shows the \emph{lower} bound from unitarity for $s$-wave annihilation. \href{https://github.com/adam-coogan/pbhs_vs_wimps/blob/master/plot_frequentist_limits.py}{\faFileCodeO}}
    \label{fig:sv_limits}
\end{figure}

For each detection scenario in Table~\ref{tab:Detections} we show as function of WIMP mass the 95\% CL upper limit on $f_\dm^4 \sv$ in Fig.~\ref{fig:sv_limits}, where $f_\dm = \Omega_\dm / \Omega_\cdm$ is the fractional contribution of a particle species to the cosmic DM density. This allows us to compare our projections with the theoretical predictions in cases where new particles constitute only a subdominant component of DM. The colored curves show the most stringent constraint arising from gamma-ray observations at a given WIMP mass, assuming annihilation into $\bar{b} b$. For our projected limits assuming a small number of PBH detections (solid lines), point source constraints dominate at low WIMP mass, while diffuse constraints are more relevant at high mass. This can be seen as a `kink' in each of the solid lines, above which diffuse constraints dominate. For larger numbers of PBH detections (dashed lines), diffuse constraints generally dominate (see Appendix~\ref{app:gamma_ray_constraints} for a more detailed comparison of the limits).

We find that a detection of $\mathcal{O}(10)$ PBHs with any of the methods described above would rule out large ranges of standard-model extensions with stable relics at the electroweak scale. To illustrate this, we show in dark grey the envelope of the 95\% CL profile-likelihood contours for the MSSM7~\cite{Athron:2017yua} and various GUT-scale SUSY models~\cite{Athron:2017qdc} obtained by the GAMBIT collaboration. These contours show the \textit{zero-velocity limit} of the annihilation cross section $\sv$ and so can be directly compared with our projected gamma-ray constraints assuming $s$-wave annihilation only. In light grey, we also show contours for scalar singlet scenarios~\cite{Athron:2017kgt}. Requiring the WIMP to comprise at least 10\% of dark matter, there are no models in the right part of the scalar singlet region below the line $f_\dm{}^4 \sv = 10^{-29}\us{cm}^3/\u{s}$, and in the leftmost sliver of the singlet contour and throughout the dark grey envelope all such models have $f_\dm{}^4 \sv > 10^{-32}\us{cm}^3/\u{s}$. 

Importantly, since $f_\dm$ scales approximately as $\langle \sigma v \rangle^{-1}$, the unitarity limit on the freeze-out annihilation cross-section $\langle \sigma v \rangle$~\cite{UnitarityGriestKamionkowski} yields, in the case of $s$-wave annihilation, a \emph{lower} bound on the rescaled annihilation cross-section today,
\begin{align}
    f_\dm{}^4 \sv \gtrsim 
    10^{-38} \us{cm}^3 \us{s}^{-1}
    \left( \frac{m_\dm}{1\us{TeV}} \right)^6 \;.
\end{align}
This is shown as the lower black dotted curve in Fig.~\ref{fig:sv_limits}. In some scenarios, we are able to exclude below this unitarity bound.
The unitarity bound can in principle be made lower in the case of $p$-wave annihilation, resonances, co-annihilation and kinematic thresholds but for the models shown this is always counterbalanced by values of $f_\chi$ that far exceed the unitarity lower bound.
Note that, although the constraints we have presented do not apply to, e.g., super-WIMP DM candidates like the gravitino~\cite{Gomez-Vargas:2016ocf,Steffen:2006hw}, they do affect asymmetric DM models~\cite{Petraki:2013wwa,Zurek:2013wia}, where the symmetric component is reduced via freeze-out. The detection of PBHs therefore places strong constraints on the identity of the particles which make up the dominant fraction of DM.

\paragraph*{Conclusions.}
Upcoming gravitational wave campaigns and radio surveys have the potential to unequivocally detect primordial black holes and therefore constrain their cosmic abundance. In this case, we have shown that even a small number of detections would place stringent constraints on self-annihilating Dark Matter. These constraints would be many orders of magnitude stronger than currents constraints from gamma-ray telescopes \cite{Abramowski:2011hc,2012ApJ...761...91A,Ackermann:2015zua,Abdallah:2018qtu} and even a single PBH detection would completely rule out thermal WIMPs which constitute a substantial fraction of the Universe's Dark Matter. These scenarios thus rule out large regions of parameter space for proposed models of new physics at the weak scale (even those with sub-dominant WIMPs), making a compelling case for the search for primordial black holes now and in the near future.

\paragraph*{Acknowledgements.} 
Where necessary, we have used the publicly available WebPlotDigitizer \cite{WebPlotDigitizer} to digitise plots.
This work was carried out on the Dutch national e-infrastructure with the support of SURF Cooperative.
BJK acknowledges funding from the Netherlands
Organization for Scientific Research (NWO) through the VIDI research program ``Probing the Genesis of Dark Matter'' (680-47-532). 
DG has received financial support through the Postdoctoral Junior Leader Fellowship Programme from la Caixa Banking Foundation (grant n.~LCF/BQ/LI18/11630014).
We thank Christian Byrnes for sharing preliminary results from~\cite{Adamek:2019gns}.

\bibliography{refs}

\begin{thebibliography}{83}%
\makeatletter
\providecommand \@ifxundefined [1]{%
 \@ifx{#1\undefined}
}%
\providecommand \@ifnum [1]{%
 \ifnum #1\expandafter \@firstoftwo
 \else \expandafter \@secondoftwo
 \fi
}%
\providecommand \@ifx [1]{%
 \ifx #1\expandafter \@firstoftwo
 \else \expandafter \@secondoftwo
 \fi
}%
\providecommand \natexlab [1]{#1}%
\providecommand \enquote  [1]{``#1''}%
\providecommand \bibnamefont  [1]{#1}%
\providecommand \bibfnamefont [1]{#1}%
\providecommand \citenamefont [1]{#1}%
\providecommand \href@noop [0]{\@secondoftwo}%
\providecommand \href [0]{\begingroup \@sanitize@url \@href}%
\providecommand \@href[1]{\@@startlink{#1}\@@href}%
\providecommand \@@href[1]{\endgroup#1\@@endlink}%
\providecommand \@sanitize@url [0]{\catcode `\\12\catcode `\$12\catcode
  `\&12\catcode `\#12\catcode `\^12\catcode `\_12\catcode `\%12\relax}%
\providecommand \@@startlink[1]{}%
\providecommand \@@endlink[0]{}%
\providecommand \url  [0]{\begingroup\@sanitize@url \@url }%
\providecommand \@url [1]{\endgroup\@href {#1}{\urlprefix }}%
\providecommand \urlprefix  [0]{URL }%
\providecommand \Eprint [0]{\href }%
\providecommand \doibase [0]{http://dx.doi.org/}%
\providecommand \selectlanguage [0]{\@gobble}%
\providecommand \bibinfo  [0]{\@secondoftwo}%
\providecommand \bibfield  [0]{\@secondoftwo}%
\providecommand \translation [1]{[#1]}%
\providecommand \BibitemOpen [0]{}%
\providecommand \bibitemStop [0]{}%
\providecommand \bibitemNoStop [0]{.\EOS\space}%
\providecommand \EOS [0]{\spacefactor3000\relax}%
\providecommand \BibitemShut  [1]{\csname bibitem#1\endcsname}%
\let\auto@bib@innerbib\@empty
\bibitem [{\citenamefont {Gondolo}\ and\ \citenamefont
  {Silk}(1999)}]{Gondolo:1999ef}%
  \BibitemOpen
  \bibfield  {author} {\bibinfo {author} {\bibfnamefont {P.}~\bibnamefont
  {Gondolo}}\ and\ \bibinfo {author} {\bibfnamefont {J.}~\bibnamefont {Silk}},\
  }\href {\doibase 10.1103/PhysRevLett.83.1719} {\bibfield  {journal} {\bibinfo
   {journal} {Phys. Rev. Lett.}\ }\textbf {\bibinfo {volume} {83}},\ \bibinfo
  {pages} {1719} (\bibinfo {year} {1999})},\ \Eprint
  {http://arxiv.org/abs/astro-ph/9906391} {arXiv:astro-ph/9906391 [astro-ph]}
  \BibitemShut {NoStop}%
\bibitem [{\citenamefont {Gondolo}(2000)}]{Gondolo:2000pn}%
  \BibitemOpen
  \bibfield  {author} {\bibinfo {author} {\bibfnamefont {P.}~\bibnamefont
  {Gondolo}},\ }\href {\doibase 10.1016/S0370-2693(00)01218-1} {\bibfield
  {journal} {\bibinfo  {journal} {Phys. Lett.}\ }\textbf {\bibinfo {volume}
  {B494}},\ \bibinfo {pages} {181} (\bibinfo {year} {2000})},\ \Eprint
  {http://arxiv.org/abs/hep-ph/0002226} {arXiv:hep-ph/0002226 [hep-ph]}
  \BibitemShut {NoStop}%
\bibitem [{\citenamefont {Ullio}\ \emph {et~al.}(2001)\citenamefont {Ullio},
  \citenamefont {Zhao},\ and\ \citenamefont {Kamionkowski}}]{Ullio:2001fb}%
  \BibitemOpen
  \bibfield  {author} {\bibinfo {author} {\bibfnamefont {P.}~\bibnamefont
  {Ullio}}, \bibinfo {author} {\bibfnamefont {H.}~\bibnamefont {Zhao}}, \ and\
  \bibinfo {author} {\bibfnamefont {M.}~\bibnamefont {Kamionkowski}},\ }\href
  {\doibase 10.1103/PhysRevD.64.043504} {\bibfield  {journal} {\bibinfo
  {journal} {Phys. Rev.}\ }\textbf {\bibinfo {volume} {D64}},\ \bibinfo {pages}
  {043504} (\bibinfo {year} {2001})},\ \Eprint
  {http://arxiv.org/abs/astro-ph/0101481} {arXiv:astro-ph/0101481 [astro-ph]}
  \BibitemShut {NoStop}%
\bibitem [{\citenamefont {Bertone}\ \emph {et~al.}(2001)\citenamefont
  {Bertone}, \citenamefont {Sigl},\ and\ \citenamefont
  {Silk}}]{Bertone:2001jv}%
  \BibitemOpen
  \bibfield  {author} {\bibinfo {author} {\bibfnamefont {G.}~\bibnamefont
  {Bertone}}, \bibinfo {author} {\bibfnamefont {G.}~\bibnamefont {Sigl}}, \
  and\ \bibinfo {author} {\bibfnamefont {J.}~\bibnamefont {Silk}},\ }\href
  {\doibase 10.1046/j.1365-8711.2001.04650.x} {\bibfield  {journal} {\bibinfo
  {journal} {Mon. Not. Roy. Astron. Soc.}\ }\textbf {\bibinfo {volume} {326}},\
  \bibinfo {pages} {799} (\bibinfo {year} {2001})},\ \Eprint
  {http://arxiv.org/abs/astro-ph/0101134} {arXiv:astro-ph/0101134 [astro-ph]}
  \BibitemShut {NoStop}%
\bibitem [{\citenamefont {Merritt}\ \emph {et~al.}(2002)\citenamefont
  {Merritt}, \citenamefont {Milosavljevic}, \citenamefont {Verde},\ and\
  \citenamefont {Jimenez}}]{Merritt:2002vj}%
  \BibitemOpen
  \bibfield  {author} {\bibinfo {author} {\bibfnamefont {D.}~\bibnamefont
  {Merritt}}, \bibinfo {author} {\bibfnamefont {M.}~\bibnamefont
  {Milosavljevic}}, \bibinfo {author} {\bibfnamefont {L.}~\bibnamefont
  {Verde}}, \ and\ \bibinfo {author} {\bibfnamefont {R.}~\bibnamefont
  {Jimenez}},\ }\href {\doibase 10.1103/PhysRevLett.88.191301} {\bibfield
  {journal} {\bibinfo  {journal} {Phys. Rev. Lett.}\ }\textbf {\bibinfo
  {volume} {88}},\ \bibinfo {pages} {191301} (\bibinfo {year} {2002})},\
  \Eprint {http://arxiv.org/abs/astro-ph/0201376} {arXiv:astro-ph/0201376
  [astro-ph]} \BibitemShut {NoStop}%
\bibitem [{\citenamefont {Bertone}\ and\ \citenamefont
  {Merritt}(2005)}]{Bertone:2005hw}%
  \BibitemOpen
  \bibfield  {author} {\bibinfo {author} {\bibfnamefont {G.}~\bibnamefont
  {Bertone}}\ and\ \bibinfo {author} {\bibfnamefont {D.}~\bibnamefont
  {Merritt}},\ }\href {\doibase 10.1103/PhysRevD.72.103502} {\bibfield
  {journal} {\bibinfo  {journal} {Phys. Rev.}\ }\textbf {\bibinfo {volume}
  {D72}},\ \bibinfo {pages} {103502} (\bibinfo {year} {2005})},\ \Eprint
  {http://arxiv.org/abs/astro-ph/0501555} {arXiv:astro-ph/0501555 [astro-ph]}
  \BibitemShut {NoStop}%
\bibitem [{\citenamefont {Merritt}\ \emph {et~al.}(2007)\citenamefont
  {Merritt}, \citenamefont {Harfst},\ and\ \citenamefont
  {Bertone}}]{Merritt:2006mt}%
  \BibitemOpen
  \bibfield  {author} {\bibinfo {author} {\bibfnamefont {D.}~\bibnamefont
  {Merritt}}, \bibinfo {author} {\bibfnamefont {S.}~\bibnamefont {Harfst}}, \
  and\ \bibinfo {author} {\bibfnamefont {G.}~\bibnamefont {Bertone}},\ }\href
  {\doibase 10.1103/PhysRevD.75.043517} {\bibfield  {journal} {\bibinfo
  {journal} {Phys. Rev.}\ }\textbf {\bibinfo {volume} {D75}},\ \bibinfo {pages}
  {043517} (\bibinfo {year} {2007})},\ \Eprint
  {http://arxiv.org/abs/astro-ph/0610425} {arXiv:astro-ph/0610425 [astro-ph]}
  \BibitemShut {NoStop}%
\bibitem [{\citenamefont {Bertone}\ \emph {et~al.}(2005)\citenamefont
  {Bertone}, \citenamefont {Zentner},\ and\ \citenamefont
  {Silk}}]{Bertone:2005xz}%
  \BibitemOpen
  \bibfield  {author} {\bibinfo {author} {\bibfnamefont {G.}~\bibnamefont
  {Bertone}}, \bibinfo {author} {\bibfnamefont {A.~R.}\ \bibnamefont
  {Zentner}}, \ and\ \bibinfo {author} {\bibfnamefont {J.}~\bibnamefont
  {Silk}},\ }\href {\doibase 10.1103/PhysRevD.72.103517} {\bibfield  {journal}
  {\bibinfo  {journal} {Phys. Rev.}\ }\textbf {\bibinfo {volume} {D72}},\
  \bibinfo {pages} {103517} (\bibinfo {year} {2005})},\ \Eprint
  {http://arxiv.org/abs/astro-ph/0509565} {arXiv:astro-ph/0509565 [astro-ph]}
  \BibitemShut {NoStop}%
\bibitem [{\citenamefont {Zhao}\ and\ \citenamefont
  {Silk}(2005)}]{Zhao:2005zr}%
  \BibitemOpen
  \bibfield  {author} {\bibinfo {author} {\bibfnamefont {H.-S.}\ \bibnamefont
  {Zhao}}\ and\ \bibinfo {author} {\bibfnamefont {J.}~\bibnamefont {Silk}},\
  }\href {\doibase 10.1103/PhysRevLett.95.011301} {\bibfield  {journal}
  {\bibinfo  {journal} {Phys. Rev. Lett.}\ }\textbf {\bibinfo {volume} {95}},\
  \bibinfo {pages} {011301} (\bibinfo {year} {2005})},\ \Eprint
  {http://arxiv.org/abs/astro-ph/0501625} {arXiv:astro-ph/0501625 [astro-ph]}
  \BibitemShut {NoStop}%
\bibitem [{\citenamefont {Bringmann}\ \emph {et~al.}(2009)\citenamefont
  {Bringmann}, \citenamefont {Lavalle},\ and\ \citenamefont
  {Salati}}]{Bringmann:2009ip}%
  \BibitemOpen
  \bibfield  {author} {\bibinfo {author} {\bibfnamefont {T.}~\bibnamefont
  {Bringmann}}, \bibinfo {author} {\bibfnamefont {J.}~\bibnamefont {Lavalle}},
  \ and\ \bibinfo {author} {\bibfnamefont {P.}~\bibnamefont {Salati}},\ }\href
  {\doibase 10.1103/PhysRevLett.103.161301} {\bibfield  {journal} {\bibinfo
  {journal} {Phys. Rev. Lett.}\ }\textbf {\bibinfo {volume} {103}},\ \bibinfo
  {pages} {161301} (\bibinfo {year} {2009})},\ \Eprint
  {http://arxiv.org/abs/0902.3665} {arXiv:0902.3665 [astro-ph.CO]} \BibitemShut
  {NoStop}%
\bibitem [{\citenamefont {Hawking}(1971)}]{Hawking:1971ei}%
  \BibitemOpen
  \bibfield  {author} {\bibinfo {author} {\bibfnamefont {S.}~\bibnamefont
  {Hawking}},\ }\href@noop {} {\bibfield  {journal} {\bibinfo  {journal} {Mon.
  Not. Roy. Astron. Soc.}\ }\textbf {\bibinfo {volume} {152}},\ \bibinfo
  {pages} {75} (\bibinfo {year} {1971})}\BibitemShut {NoStop}%
\bibitem [{\citenamefont {Carr}\ and\ \citenamefont
  {Hawking}(1974)}]{Carr:1974nx}%
  \BibitemOpen
  \bibfield  {author} {\bibinfo {author} {\bibfnamefont {B.~J.}\ \bibnamefont
  {Carr}}\ and\ \bibinfo {author} {\bibfnamefont {S.~W.}\ \bibnamefont
  {Hawking}},\ }\href@noop {} {\bibfield  {journal} {\bibinfo  {journal} {Mon.
  Not. Roy. Astron. Soc.}\ }\textbf {\bibinfo {volume} {168}},\ \bibinfo
  {pages} {399} (\bibinfo {year} {1974})}\BibitemShut {NoStop}%
\bibitem [{\citenamefont {Lacki}\ and\ \citenamefont
  {Beacom}(2010)}]{Lacki:2010zf}%
  \BibitemOpen
  \bibfield  {author} {\bibinfo {author} {\bibfnamefont {B.~C.}\ \bibnamefont
  {Lacki}}\ and\ \bibinfo {author} {\bibfnamefont {J.~F.}\ \bibnamefont
  {Beacom}},\ }\href {\doibase 10.1088/2041-8205/720/1/L67} {\bibfield
  {journal} {\bibinfo  {journal} {Astrophys. J.}\ }\textbf {\bibinfo {volume}
  {720}},\ \bibinfo {pages} {L67} (\bibinfo {year} {2010})},\ \Eprint
  {http://arxiv.org/abs/1003.3466} {arXiv:1003.3466 [astro-ph.CO]} \BibitemShut
  {NoStop}%
\bibitem [{\citenamefont {Eroshenko}(2016)}]{Eroshenko:2016yve}%
  \BibitemOpen
  \bibfield  {author} {\bibinfo {author} {\bibfnamefont {{\relax Yu}.~N.}\
  \bibnamefont {Eroshenko}},\ }\href {\doibase 10.1134/S1063773716060013}
  {\bibfield  {journal} {\bibinfo  {journal} {Astron. Lett.}\ }\textbf
  {\bibinfo {volume} {42}},\ \bibinfo {pages} {347} (\bibinfo {year} {2016})},\
  \bibinfo {note} {[Pisma Astron. Zh.42,no.6,359(2016)]},\ \Eprint
  {http://arxiv.org/abs/1607.00612} {arXiv:1607.00612 [astro-ph.HE]}
  \BibitemShut {NoStop}%
\bibitem [{\citenamefont {Boucenna}\ \emph {et~al.}(2018)\citenamefont
  {Boucenna}, \citenamefont {Kuhnel}, \citenamefont {Ohlsson},\ and\
  \citenamefont {Visinelli}}]{Boucenna:2017ghj}%
  \BibitemOpen
  \bibfield  {author} {\bibinfo {author} {\bibfnamefont {S.~M.}\ \bibnamefont
  {Boucenna}}, \bibinfo {author} {\bibfnamefont {F.}~\bibnamefont {Kuhnel}},
  \bibinfo {author} {\bibfnamefont {T.}~\bibnamefont {Ohlsson}}, \ and\
  \bibinfo {author} {\bibfnamefont {L.}~\bibnamefont {Visinelli}},\ }\href
  {\doibase 10.1088/1475-7516/2018/07/003} {\bibfield  {journal} {\bibinfo
  {journal} {JCAP}\ }\textbf {\bibinfo {volume} {1807}},\ \bibinfo {pages}
  {003} (\bibinfo {year} {2018})},\ \Eprint {http://arxiv.org/abs/1712.06383}
  {arXiv:1712.06383 [hep-ph]} \BibitemShut {NoStop}%
\bibitem [{\citenamefont {Adamek}\ \emph {et~al.}(2019)\citenamefont {Adamek},
  \citenamefont {Byrnes}, \citenamefont {Gosenca},\ and\ \citenamefont
  {Hotchkiss}}]{Adamek:2019gns}%
  \BibitemOpen
  \bibfield  {author} {\bibinfo {author} {\bibfnamefont {J.}~\bibnamefont
  {Adamek}}, \bibinfo {author} {\bibfnamefont {C.~T.}\ \bibnamefont {Byrnes}},
  \bibinfo {author} {\bibfnamefont {M.}~\bibnamefont {Gosenca}}, \ and\
  \bibinfo {author} {\bibfnamefont {S.}~\bibnamefont {Hotchkiss}},\ }\href@noop
  {} {\  (\bibinfo {year} {2019})},\ \Eprint {http://arxiv.org/abs/1901.08528}
  {arXiv:1901.08528 [astro-ph.CO]} \BibitemShut {NoStop}%
\bibitem [{\citenamefont {Aasi}\ \emph {et~al.}(2013)\citenamefont {Aasi} \emph
  {et~al.}}]{Aasi:2013jjl}%
  \BibitemOpen
  \bibfield  {author} {\bibinfo {author} {\bibfnamefont {J.}~\bibnamefont
  {Aasi}} \emph {et~al.} (\bibinfo {collaboration} {LIGO Scientific, VIRGO}),\
  }\href {\doibase 10.1103/PhysRevD.88.062001} {\bibfield  {journal} {\bibinfo
  {journal} {Phys. Rev.}\ }\textbf {\bibinfo {volume} {D88}},\ \bibinfo {pages}
  {062001} (\bibinfo {year} {2013})},\ \Eprint {http://arxiv.org/abs/1304.1775}
  {arXiv:1304.1775 [gr-qc]} \BibitemShut {NoStop}%
\bibitem [{\citenamefont {Slany}\ and\ \citenamefont
  {Stuchlik}(2008)}]{Slany:2008pm}%
  \BibitemOpen
  \bibfield  {author} {\bibinfo {author} {\bibfnamefont {P.}~\bibnamefont
  {Slany}}\ and\ \bibinfo {author} {\bibfnamefont {Z.}~\bibnamefont
  {Stuchlik}},\ }\href {\doibase 10.1051/0004-6361:200810334} {\bibfield
  {journal} {\bibinfo  {journal} {Astron. Astrophys.}\ }\textbf {\bibinfo
  {volume} {492}},\ \bibinfo {pages} {319} (\bibinfo {year} {2008})},\ \Eprint
  {http://arxiv.org/abs/0810.0237} {arXiv:0810.0237 [astro-ph]} \BibitemShut
  {NoStop}%
\bibitem [{\citenamefont {Shaposhnikov}\ and\ \citenamefont
  {Titarchuk}(2009)}]{Shaposhnikov:2009dc}%
  \BibitemOpen
  \bibfield  {author} {\bibinfo {author} {\bibfnamefont {N.}~\bibnamefont
  {Shaposhnikov}}\ and\ \bibinfo {author} {\bibfnamefont {L.}~\bibnamefont
  {Titarchuk}},\ }\href {\doibase 10.1088/0004-637X/699/1/453} {\bibfield
  {journal} {\bibinfo  {journal} {Astrophys. J.}\ }\textbf {\bibinfo {volume}
  {699}},\ \bibinfo {pages} {453} (\bibinfo {year} {2009})},\ \Eprint
  {http://arxiv.org/abs/0902.2852} {arXiv:0902.2852 [astro-ph.HE]} \BibitemShut
  {NoStop}%
\bibitem [{\citenamefont
  {{Chandrasekhar}}(1931{\natexlab{a}})}]{1931ApJ....74...81C}%
  \BibitemOpen
  \bibfield  {author} {\bibinfo {author} {\bibfnamefont {S.}~\bibnamefont
  {{Chandrasekhar}}},\ }\href {\doibase 10.1086/143324} {\bibfield  {journal}
  {\bibinfo  {journal} {{Astrophys.~J.}}\ }\textbf {\bibinfo {volume} {74}},\
  \bibinfo {pages} {81} (\bibinfo {year} {1931}{\natexlab{a}})}\BibitemShut
  {NoStop}%
\bibitem [{\citenamefont
  {{Chandrasekhar}}(1931{\natexlab{b}})}]{1931MNRAS..91..456C}%
  \BibitemOpen
  \bibfield  {author} {\bibinfo {author} {\bibfnamefont {S.}~\bibnamefont
  {{Chandrasekhar}}},\ }\href {\doibase 10.1093/mnras/91.5.456} {\bibfield
  {journal} {\bibinfo  {journal} {{Mon.~Not.~Roy.~Astron.~Soc.}}\ }\textbf
  {\bibinfo {volume} {91}},\ \bibinfo {pages} {456} (\bibinfo {year}
  {1931}{\natexlab{b}})}\BibitemShut {NoStop}%
\bibitem [{\citenamefont {{Chandrasekhar}}(1935)}]{1935MNRAS..95..207C}%
  \BibitemOpen
  \bibfield  {author} {\bibinfo {author} {\bibfnamefont {S.}~\bibnamefont
  {{Chandrasekhar}}},\ }\href {\doibase 10.1093/mnras/95.3.207} {\bibfield
  {journal} {\bibinfo  {journal} {{Mon.~Not.~Roy.~Astron.~Soc.}}\ }\textbf
  {\bibinfo {volume} {95}},\ \bibinfo {pages} {207} (\bibinfo {year}
  {1935})}\BibitemShut {NoStop}%
\bibitem [{\citenamefont {Shandera}\ \emph {et~al.}(2018)\citenamefont
  {Shandera}, \citenamefont {Jeong},\ and\ \citenamefont
  {Gebhardt}}]{Shandera:2018xkn}%
  \BibitemOpen
  \bibfield  {author} {\bibinfo {author} {\bibfnamefont {S.}~\bibnamefont
  {Shandera}}, \bibinfo {author} {\bibfnamefont {D.}~\bibnamefont {Jeong}}, \
  and\ \bibinfo {author} {\bibfnamefont {H.~S.~G.}\ \bibnamefont {Gebhardt}},\
  }\href {\doibase 10.1103/PhysRevLett.120.241102} {\bibfield  {journal}
  {\bibinfo  {journal} {Phys. Rev. Lett.}\ }\textbf {\bibinfo {volume} {120}},\
  \bibinfo {pages} {241102} (\bibinfo {year} {2018})},\ \Eprint
  {http://arxiv.org/abs/1802.08206} {arXiv:1802.08206 [astro-ph.CO]}
  \BibitemShut {NoStop}%
\bibitem [{\citenamefont {Kouvaris}\ \emph {et~al.}(2018)\citenamefont
  {Kouvaris}, \citenamefont {Tinyakov},\ and\ \citenamefont
  {Tytgat}}]{Kouvaris:2018wnh}%
  \BibitemOpen
  \bibfield  {author} {\bibinfo {author} {\bibfnamefont {C.}~\bibnamefont
  {Kouvaris}}, \bibinfo {author} {\bibfnamefont {P.}~\bibnamefont {Tinyakov}},
  \ and\ \bibinfo {author} {\bibfnamefont {M.~H.~G.}\ \bibnamefont {Tytgat}},\
  }\href {\doibase 10.1103/PhysRevLett.121.221102} {\bibfield  {journal}
  {\bibinfo  {journal} {Phys. Rev. Lett.}\ }\textbf {\bibinfo {volume} {121}},\
  \bibinfo {pages} {221102} (\bibinfo {year} {2018})},\ \Eprint
  {http://arxiv.org/abs/1804.06740} {arXiv:1804.06740 [astro-ph.HE]}
  \BibitemShut {NoStop}%
\bibitem [{\citenamefont {Hild}\ \emph {et~al.}(2011)\citenamefont {Hild} \emph
  {et~al.}}]{Hild:2010id}%
  \BibitemOpen
  \bibfield  {author} {\bibinfo {author} {\bibfnamefont {S.}~\bibnamefont
  {Hild}} \emph {et~al.},\ }\href {\doibase 10.1088/0264-9381/28/9/094013}
  {\bibfield  {journal} {\bibinfo  {journal} {Class. Quant. Grav.}\ }\textbf
  {\bibinfo {volume} {28}},\ \bibinfo {pages} {094013} (\bibinfo {year}
  {2011})},\ \Eprint {http://arxiv.org/abs/1012.0908} {arXiv:1012.0908 [gr-qc]}
  \BibitemShut {NoStop}%
\bibitem [{\citenamefont {Sathyaprakash}\ \emph {et~al.}(2012)\citenamefont
  {Sathyaprakash} \emph {et~al.}}]{Sathyaprakash:2012jk}%
  \BibitemOpen
  \bibfield  {author} {\bibinfo {author} {\bibfnamefont {B.}~\bibnamefont
  {Sathyaprakash}} \emph {et~al.},\ }\bibfield  {booktitle} {\emph {\bibinfo
  {booktitle} {{Gravitational waves. Numerical relativity - data analysis.
  Proceedings, 9th Edoardo Amaldi Conference, Amaldi 9, and meeting, NRDA 2011,
  Cardiff, UK, July 10-15, 2011}}},\ }\href {\doibase
  10.1088/0264-9381/29/12/124013, 10.1088/0264-9381/30/7/079501} {\bibfield
  {journal} {\bibinfo  {journal} {Class. Quant. Grav.}\ }\textbf {\bibinfo
  {volume} {29}},\ \bibinfo {pages} {124013} (\bibinfo {year} {2012})},\
  \bibinfo {note} {[Erratum: Class. Quant. Grav.30,079501(2013)]},\ \Eprint
  {http://arxiv.org/abs/1206.0331} {arXiv:1206.0331 [gr-qc]} \BibitemShut
  {NoStop}%
\bibitem [{\citenamefont {Mapelli}\ \emph {et~al.}(2019)\citenamefont
  {Mapelli}, \citenamefont {Giacobbo}, \citenamefont {Santoliquido},\ and\
  \citenamefont {Artale}}]{Mapelli:2019bnp}%
  \BibitemOpen
  \bibfield  {author} {\bibinfo {author} {\bibfnamefont {M.}~\bibnamefont
  {Mapelli}}, \bibinfo {author} {\bibfnamefont {N.}~\bibnamefont {Giacobbo}},
  \bibinfo {author} {\bibfnamefont {F.}~\bibnamefont {Santoliquido}}, \ and\
  \bibinfo {author} {\bibfnamefont {M.~C.}\ \bibnamefont {Artale}},\
  }\href@noop {} {\  (\bibinfo {year} {2019})},\ \Eprint
  {http://arxiv.org/abs/1902.01419} {arXiv:1902.01419 [astro-ph.HE]}
  \BibitemShut {NoStop}%
\bibitem [{\citenamefont {Koushiappas}\ and\ \citenamefont
  {Loeb}(2017)}]{Koushiappas:2017kqm}%
  \BibitemOpen
  \bibfield  {author} {\bibinfo {author} {\bibfnamefont {S.~M.}\ \bibnamefont
  {Koushiappas}}\ and\ \bibinfo {author} {\bibfnamefont {A.}~\bibnamefont
  {Loeb}},\ }\href {\doibase 10.1103/PhysRevLett.119.221104} {\bibfield
  {journal} {\bibinfo  {journal} {Phys. Rev. Lett.}\ }\textbf {\bibinfo
  {volume} {119}},\ \bibinfo {pages} {221104} (\bibinfo {year} {2017})},\
  \Eprint {http://arxiv.org/abs/1708.07380} {arXiv:1708.07380 [astro-ph.CO]}
  \BibitemShut {NoStop}%
\bibitem [{\citenamefont {Chen}\ and\ \citenamefont
  {Huang}(2019)}]{Chen:2019irf}%
  \BibitemOpen
  \bibfield  {author} {\bibinfo {author} {\bibfnamefont {Z.-C.}\ \bibnamefont
  {Chen}}\ and\ \bibinfo {author} {\bibfnamefont {Q.-G.}\ \bibnamefont
  {Huang}},\ }\href@noop {} {\  (\bibinfo {year} {2019})},\ \Eprint
  {http://arxiv.org/abs/1904.02396} {arXiv:1904.02396 [astro-ph.CO]}
  \BibitemShut {NoStop}%
\bibitem [{\citenamefont {Ali-Haïmoud}\ \emph {et~al.}(2017)\citenamefont
  {Ali-Haïmoud}, \citenamefont {Kovetz},\ and\ \citenamefont
  {Kamionkowski}}]{Ali-Haimoud:2017rtz}%
  \BibitemOpen
  \bibfield  {author} {\bibinfo {author} {\bibfnamefont {Y.}~\bibnamefont
  {Ali-Haïmoud}}, \bibinfo {author} {\bibfnamefont {E.~D.}\ \bibnamefont
  {Kovetz}}, \ and\ \bibinfo {author} {\bibfnamefont {M.}~\bibnamefont
  {Kamionkowski}},\ }\href {\doibase 10.1103/PhysRevD.96.123523} {\bibfield
  {journal} {\bibinfo  {journal} {Phys. Rev.}\ }\textbf {\bibinfo {volume}
  {D96}},\ \bibinfo {pages} {123523} (\bibinfo {year} {2017})},\ \Eprint
  {http://arxiv.org/abs/1709.06576} {arXiv:1709.06576 [astro-ph.CO]}
  \BibitemShut {NoStop}%
\bibitem [{\citenamefont {Kavanagh}\ \emph {et~al.}(2018)\citenamefont
  {Kavanagh}, \citenamefont {Gaggero},\ and\ \citenamefont
  {Bertone}}]{Kavanagh:2018ggo}%
  \BibitemOpen
  \bibfield  {author} {\bibinfo {author} {\bibfnamefont {B.~J.}\ \bibnamefont
  {Kavanagh}}, \bibinfo {author} {\bibfnamefont {D.}~\bibnamefont {Gaggero}}, \
  and\ \bibinfo {author} {\bibfnamefont {G.}~\bibnamefont {Bertone}},\ }\href
  {\doibase 10.1103/PhysRevD.98.023536} {\bibfield  {journal} {\bibinfo
  {journal} {Phys. Rev.}\ }\textbf {\bibinfo {volume} {D98}},\ \bibinfo {pages}
  {023536} (\bibinfo {year} {2018})},\ \Eprint
  {http://arxiv.org/abs/1805.09034} {arXiv:1805.09034 [astro-ph.CO]}
  \BibitemShut {NoStop}%
\bibitem [{\citenamefont {Hayasaki}\ \emph {et~al.}(2016)\citenamefont
  {Hayasaki}, \citenamefont {Takahashi}, \citenamefont {Sendouda},\ and\
  \citenamefont {Nagataki}}]{Hayasaki:2009ug}%
  \BibitemOpen
  \bibfield  {author} {\bibinfo {author} {\bibfnamefont {K.}~\bibnamefont
  {Hayasaki}}, \bibinfo {author} {\bibfnamefont {K.}~\bibnamefont {Takahashi}},
  \bibinfo {author} {\bibfnamefont {Y.}~\bibnamefont {Sendouda}}, \ and\
  \bibinfo {author} {\bibfnamefont {S.}~\bibnamefont {Nagataki}},\ }\href
  {\doibase 10.1093/pasj/psw065} {\bibfield  {journal} {\bibinfo  {journal}
  {Publ. Astron. Soc. Jap.}\ }\textbf {\bibinfo {volume} {68}},\ \bibinfo
  {pages} {66} (\bibinfo {year} {2016})},\ \Eprint
  {http://arxiv.org/abs/0909.1738} {arXiv:0909.1738 [astro-ph.CO]} \BibitemShut
  {NoStop}%
\bibitem [{\citenamefont {Raidal}\ \emph {et~al.}(2017)\citenamefont {Raidal},
  \citenamefont {Vaskonen},\ and\ \citenamefont {Veermäe}}]{Raidal:2017mfl}%
  \BibitemOpen
  \bibfield  {author} {\bibinfo {author} {\bibfnamefont {M.}~\bibnamefont
  {Raidal}}, \bibinfo {author} {\bibfnamefont {V.}~\bibnamefont {Vaskonen}}, \
  and\ \bibinfo {author} {\bibfnamefont {H.}~\bibnamefont {Veermäe}},\ }\href
  {\doibase 10.1088/1475-7516/2017/09/037} {\bibfield  {journal} {\bibinfo
  {journal} {JCAP}\ }\textbf {\bibinfo {volume} {1709}},\ \bibinfo {pages}
  {037} (\bibinfo {year} {2017})},\ \Eprint {http://arxiv.org/abs/1707.01480}
  {arXiv:1707.01480 [astro-ph.CO]} \BibitemShut {NoStop}%
\bibitem [{\citenamefont {Chen}\ and\ \citenamefont
  {Huang}(2018)}]{Chen:2018czv}%
  \BibitemOpen
  \bibfield  {author} {\bibinfo {author} {\bibfnamefont {Z.-C.}\ \bibnamefont
  {Chen}}\ and\ \bibinfo {author} {\bibfnamefont {Q.-G.}\ \bibnamefont
  {Huang}},\ }\href@noop {} {\  (\bibinfo {year} {2018})},\ \Eprint
  {http://arxiv.org/abs/1801.10327} {arXiv:1801.10327 [astro-ph.CO]}
  \BibitemShut {NoStop}%
\bibitem [{\citenamefont {Ballesteros}\ \emph {et~al.}(2018)\citenamefont
  {Ballesteros}, \citenamefont {Serpico},\ and\ \citenamefont
  {Taoso}}]{Ballesteros:2018swv}%
  \BibitemOpen
  \bibfield  {author} {\bibinfo {author} {\bibfnamefont {G.}~\bibnamefont
  {Ballesteros}}, \bibinfo {author} {\bibfnamefont {P.~D.}\ \bibnamefont
  {Serpico}}, \ and\ \bibinfo {author} {\bibfnamefont {M.}~\bibnamefont
  {Taoso}},\ }\href {\doibase 10.1088/1475-7516/2018/10/043} {\bibfield
  {journal} {\bibinfo  {journal} {JCAP}\ }\textbf {\bibinfo {volume} {1810}},\
  \bibinfo {pages} {043} (\bibinfo {year} {2018})},\ \Eprint
  {http://arxiv.org/abs/1807.02084} {arXiv:1807.02084 [astro-ph.CO]}
  \BibitemShut {NoStop}%
\bibitem [{\citenamefont {Belotsky}\ \emph {et~al.}(2019)\citenamefont
  {Belotsky}, \citenamefont {Dokuchaev}, \citenamefont {Eroshenko},
  \citenamefont {Esipova}, \citenamefont {Khlopov}, \citenamefont {Khromykh},
  \citenamefont {Kirillov}, \citenamefont {Nikulin}, \citenamefont {Rubin},\
  and\ \citenamefont {Svadkovsky}}]{Belotsky:2018wph}%
  \BibitemOpen
  \bibfield  {author} {\bibinfo {author} {\bibfnamefont {K.~M.}\ \bibnamefont
  {Belotsky}}, \bibinfo {author} {\bibfnamefont {V.~I.}\ \bibnamefont
  {Dokuchaev}}, \bibinfo {author} {\bibfnamefont {Y.~N.}\ \bibnamefont
  {Eroshenko}}, \bibinfo {author} {\bibfnamefont {E.~A.}\ \bibnamefont
  {Esipova}}, \bibinfo {author} {\bibfnamefont {M.~{\relax Yu}.}\ \bibnamefont
  {Khlopov}}, \bibinfo {author} {\bibfnamefont {L.~A.}\ \bibnamefont
  {Khromykh}}, \bibinfo {author} {\bibfnamefont {A.~A.}\ \bibnamefont
  {Kirillov}}, \bibinfo {author} {\bibfnamefont {V.~V.}\ \bibnamefont
  {Nikulin}}, \bibinfo {author} {\bibfnamefont {S.~G.}\ \bibnamefont {Rubin}},
  \ and\ \bibinfo {author} {\bibfnamefont {I.~V.}\ \bibnamefont {Svadkovsky}},\
  }\href {\doibase 10.1140/epjc/s10052-019-6741-4} {\bibfield  {journal}
  {\bibinfo  {journal} {Eur. Phys. J.}\ }\textbf {\bibinfo {volume} {C79}},\
  \bibinfo {pages} {246} (\bibinfo {year} {2019})},\ \Eprint
  {http://arxiv.org/abs/1807.06590} {arXiv:1807.06590 [astro-ph.CO]}
  \BibitemShut {NoStop}%
\bibitem [{\citenamefont {Bringmann}\ \emph {et~al.}(2019)\citenamefont
  {Bringmann}, \citenamefont {Depta}, \citenamefont {Domcke},\ and\
  \citenamefont {Schmidt-Hoberg}}]{Bringmann:2018mxj}%
  \BibitemOpen
  \bibfield  {author} {\bibinfo {author} {\bibfnamefont {T.}~\bibnamefont
  {Bringmann}}, \bibinfo {author} {\bibfnamefont {P.~F.}\ \bibnamefont
  {Depta}}, \bibinfo {author} {\bibfnamefont {V.}~\bibnamefont {Domcke}}, \
  and\ \bibinfo {author} {\bibfnamefont {K.}~\bibnamefont {Schmidt-Hoberg}},\
  }\href {\doibase 10.1103/PhysRevD.99.063532} {\bibfield  {journal} {\bibinfo
  {journal} {Phys. Rev.}\ }\textbf {\bibinfo {volume} {D99}},\ \bibinfo {pages}
  {063532} (\bibinfo {year} {2019})},\ \Eprint
  {http://arxiv.org/abs/1808.05910} {arXiv:1808.05910 [astro-ph.CO]}
  \BibitemShut {NoStop}%
\bibitem [{\citenamefont {Raidal}\ \emph {et~al.}(2019)\citenamefont {Raidal},
  \citenamefont {Spethmann}, \citenamefont {Vaskonen},\ and\ \citenamefont
  {Veermäe}}]{Raidal:2018bbj}%
  \BibitemOpen
  \bibfield  {author} {\bibinfo {author} {\bibfnamefont {M.}~\bibnamefont
  {Raidal}}, \bibinfo {author} {\bibfnamefont {C.}~\bibnamefont {Spethmann}},
  \bibinfo {author} {\bibfnamefont {V.}~\bibnamefont {Vaskonen}}, \ and\
  \bibinfo {author} {\bibfnamefont {H.}~\bibnamefont {Veermäe}},\ }\href
  {\doibase 10.1088/1475-7516/2019/02/018} {\bibfield  {journal} {\bibinfo
  {journal} {JCAP}\ }\textbf {\bibinfo {volume} {1902}},\ \bibinfo {pages}
  {018} (\bibinfo {year} {2019})},\ \Eprint {http://arxiv.org/abs/1812.01930}
  {arXiv:1812.01930 [astro-ph.CO]} \BibitemShut {NoStop}%
\bibitem [{\citenamefont {Nakamura}\ \emph {et~al.}(1997)\citenamefont
  {Nakamura}, \citenamefont {Sasaki}, \citenamefont {Tanaka},\ and\
  \citenamefont {Thorne}}]{Nakamura:1997sm}%
  \BibitemOpen
  \bibfield  {author} {\bibinfo {author} {\bibfnamefont {T.}~\bibnamefont
  {Nakamura}}, \bibinfo {author} {\bibfnamefont {M.}~\bibnamefont {Sasaki}},
  \bibinfo {author} {\bibfnamefont {T.}~\bibnamefont {Tanaka}}, \ and\ \bibinfo
  {author} {\bibfnamefont {K.~S.}\ \bibnamefont {Thorne}},\ }\href {\doibase
  10.1086/310886} {\bibfield  {journal} {\bibinfo  {journal} {Astrophys. J.}\
  }\textbf {\bibinfo {volume} {487}},\ \bibinfo {pages} {L139} (\bibinfo {year}
  {1997})},\ \Eprint {http://arxiv.org/abs/astro-ph/9708060}
  {arXiv:astro-ph/9708060 [astro-ph]} \BibitemShut {NoStop}%
\bibitem [{\citenamefont {Ioka}\ \emph {et~al.}(1998)\citenamefont {Ioka},
  \citenamefont {Chiba}, \citenamefont {Tanaka},\ and\ \citenamefont
  {Nakamura}}]{Ioka:1998nz}%
  \BibitemOpen
  \bibfield  {author} {\bibinfo {author} {\bibfnamefont {K.}~\bibnamefont
  {Ioka}}, \bibinfo {author} {\bibfnamefont {T.}~\bibnamefont {Chiba}},
  \bibinfo {author} {\bibfnamefont {T.}~\bibnamefont {Tanaka}}, \ and\ \bibinfo
  {author} {\bibfnamefont {T.}~\bibnamefont {Nakamura}},\ }\href {\doibase
  10.1103/PhysRevD.58.063003} {\bibfield  {journal} {\bibinfo  {journal} {Phys.
  Rev.}\ }\textbf {\bibinfo {volume} {D58}},\ \bibinfo {pages} {063003}
  (\bibinfo {year} {1998})},\ \Eprint {http://arxiv.org/abs/astro-ph/9807018}
  {arXiv:astro-ph/9807018 [astro-ph]} \BibitemShut {NoStop}%
\bibitem [{\citenamefont {Sasaki}\ \emph {et~al.}(2016)\citenamefont {Sasaki},
  \citenamefont {Suyama}, \citenamefont {Tanaka},\ and\ \citenamefont
  {Yokoyama}}]{Sasaki:2016jop}%
  \BibitemOpen
  \bibfield  {author} {\bibinfo {author} {\bibfnamefont {M.}~\bibnamefont
  {Sasaki}}, \bibinfo {author} {\bibfnamefont {T.}~\bibnamefont {Suyama}},
  \bibinfo {author} {\bibfnamefont {T.}~\bibnamefont {Tanaka}}, \ and\ \bibinfo
  {author} {\bibfnamefont {S.}~\bibnamefont {Yokoyama}},\ }\href {\doibase
  10.1103/PhysRevLett.121.059901, 10.1103/PhysRevLett.117.061101} {\bibfield
  {journal} {\bibinfo  {journal} {Phys. Rev. Lett.}\ }\textbf {\bibinfo
  {volume} {117}},\ \bibinfo {pages} {061101} (\bibinfo {year} {2016})},\
  \bibinfo {note} {[erratum: Phys. Rev. Lett.121,no.5,059901(2018)]},\ \Eprint
  {http://arxiv.org/abs/1603.08338} {arXiv:1603.08338 [astro-ph.CO]}
  \BibitemShut {NoStop}%
\bibitem [{\citenamefont {Peters}(1964)}]{Peters:1964zz}%
  \BibitemOpen
  \bibfield  {author} {\bibinfo {author} {\bibfnamefont {P.~C.}\ \bibnamefont
  {Peters}},\ }\href {\doibase 10.1103/PhysRev.136.B1224} {\bibfield  {journal}
  {\bibinfo  {journal} {Phys. Rev.}\ }\textbf {\bibinfo {volume} {136}},\
  \bibinfo {pages} {B1224} (\bibinfo {year} {1964})}\BibitemShut {NoStop}%
\bibitem [{\citenamefont {Abbott}\ \emph
  {et~al.}(2016{\natexlab{a}})\citenamefont {Abbott} \emph
  {et~al.}}]{Abbott:2016nhf}%
  \BibitemOpen
  \bibfield  {author} {\bibinfo {author} {\bibfnamefont {B.~P.}\ \bibnamefont
  {Abbott}} \emph {et~al.} (\bibinfo {collaboration} {Virgo, LIGO
  Scientific}),\ }\href {\doibase 10.3847/2041-8205/833/1/L1} {\bibfield
  {journal} {\bibinfo  {journal} {Astrophys. J.}\ }\textbf {\bibinfo {volume}
  {833}},\ \bibinfo {pages} {L1} (\bibinfo {year} {2016}{\natexlab{a}})},\
  \Eprint {http://arxiv.org/abs/1602.03842} {arXiv:1602.03842 [astro-ph.HE]}
  \BibitemShut {NoStop}%
\bibitem [{\citenamefont {Abbott}\ \emph
  {et~al.}(2016{\natexlab{b}})\citenamefont {Abbott} \emph
  {et~al.}}]{Abbott:2016drs}%
  \BibitemOpen
  \bibfield  {author} {\bibinfo {author} {\bibfnamefont {B.~P.}\ \bibnamefont
  {Abbott}} \emph {et~al.} (\bibinfo {collaboration} {Virgo, LIGO
  Scientific}),\ }\href {\doibase 10.3847/0067-0049/227/2/14} {\bibfield
  {journal} {\bibinfo  {journal} {Astrophys. J. Suppl.}\ }\textbf {\bibinfo
  {volume} {227}},\ \bibinfo {pages} {14} (\bibinfo {year}
  {2016}{\natexlab{b}})},\ \Eprint {http://arxiv.org/abs/1606.03939}
  {arXiv:1606.03939 [astro-ph.HE]} \BibitemShut {NoStop}%
\bibitem [{\citenamefont {Hogg}(1999)}]{Hogg:1999ad}%
  \BibitemOpen
  \bibfield  {author} {\bibinfo {author} {\bibfnamefont {D.~W.}\ \bibnamefont
  {Hogg}},\ }\href@noop {} {\  (\bibinfo {year} {1999})},\ \Eprint
  {http://arxiv.org/abs/astro-ph/9905116} {arXiv:astro-ph/9905116 [astro-ph]}
  \BibitemShut {NoStop}%
\bibitem [{\citenamefont {Aghanim}\ \emph {et~al.}(2018)\citenamefont {Aghanim}
  \emph {et~al.}}]{Aghanim:2018eyx}%
  \BibitemOpen
  \bibfield  {author} {\bibinfo {author} {\bibfnamefont {N.}~\bibnamefont
  {Aghanim}} \emph {et~al.} (\bibinfo {collaboration} {Planck}),\ }\href@noop
  {} {\  (\bibinfo {year} {2018})},\ \Eprint {http://arxiv.org/abs/1807.06209}
  {arXiv:1807.06209 [astro-ph.CO]} \BibitemShut {NoStop}%
\bibitem [{\citenamefont {Jeffreys}(1946)}]{Jeffreys}%
  \BibitemOpen
  \bibfield  {author} {\bibinfo {author} {\bibfnamefont {H.}~\bibnamefont
  {Jeffreys}},\ }\href {\doibase 10.1098/rspa.1946.0056} {\bibfield  {journal}
  {\bibinfo  {journal} {Proceedings of the Royal Society of London. Series A.
  Mathematical and Physical Sciences}\ }\textbf {\bibinfo {volume} {186}},\
  \bibinfo {pages} {453} (\bibinfo {year} {1946})}\BibitemShut {NoStop}%
\bibitem [{\citenamefont {Bayes}(1764)}]{Bayes:1764vd}%
  \BibitemOpen
  \bibfield  {author} {\bibinfo {author} {\bibfnamefont {R.}~\bibnamefont
  {Bayes}},\ }\href {\doibase 10.1098/rstl.1763.0053} {\bibfield  {journal}
  {\bibinfo  {journal} {Phil. Trans. Roy. Soc. Lond.}\ }\textbf {\bibinfo
  {volume} {53}},\ \bibinfo {pages} {370} (\bibinfo {year} {1764})}\BibitemShut
  {NoStop}%
\bibitem [{\citenamefont {Magee}\ \emph {et~al.}(2018)\citenamefont {Magee},
  \citenamefont {Deutsch}, \citenamefont {McClincy}, \citenamefont {Hanna},
  \citenamefont {Horst}, \citenamefont {Meacher}, \citenamefont {Messick},
  \citenamefont {Shandera},\ and\ \citenamefont {Wade}}]{Magee:2018opb}%
  \BibitemOpen
  \bibfield  {author} {\bibinfo {author} {\bibfnamefont {R.}~\bibnamefont
  {Magee}}, \bibinfo {author} {\bibfnamefont {A.-S.}\ \bibnamefont {Deutsch}},
  \bibinfo {author} {\bibfnamefont {P.}~\bibnamefont {McClincy}}, \bibinfo
  {author} {\bibfnamefont {C.}~\bibnamefont {Hanna}}, \bibinfo {author}
  {\bibfnamefont {C.}~\bibnamefont {Horst}}, \bibinfo {author} {\bibfnamefont
  {D.}~\bibnamefont {Meacher}}, \bibinfo {author} {\bibfnamefont
  {C.}~\bibnamefont {Messick}}, \bibinfo {author} {\bibfnamefont
  {S.}~\bibnamefont {Shandera}}, \ and\ \bibinfo {author} {\bibfnamefont
  {M.}~\bibnamefont {Wade}},\ }\href {\doibase 10.1103/PhysRevD.98.103024}
  {\bibfield  {journal} {\bibinfo  {journal} {Phys. Rev.}\ }\textbf {\bibinfo
  {volume} {D98}},\ \bibinfo {pages} {103024} (\bibinfo {year} {2018})},\
  \Eprint {http://arxiv.org/abs/1808.04772} {arXiv:1808.04772 [astro-ph.IM]}
  \BibitemShut {NoStop}%
\bibitem [{\citenamefont {Abbott}\ \emph
  {et~al.}(2018{\natexlab{a}})\citenamefont {Abbott} \emph
  {et~al.}}]{Aasi:2013wya}%
  \BibitemOpen
  \bibfield  {author} {\bibinfo {author} {\bibfnamefont {B.~P.}\ \bibnamefont
  {Abbott}} \emph {et~al.} (\bibinfo {collaboration} {KAGRA, LIGO Scientific,
  VIRGO}),\ }\href {\doibase 10.1007/s41114-018-0012-9, 10.1007/lrr-2016-1}
  {\bibfield  {journal} {\bibinfo  {journal} {Living Rev. Rel.}\ }\textbf
  {\bibinfo {volume} {21}},\ \bibinfo {pages} {3} (\bibinfo {year}
  {2018}{\natexlab{a}})},\ \Eprint {http://arxiv.org/abs/1304.0670}
  {arXiv:1304.0670 [gr-qc]} \BibitemShut {NoStop}%
\bibitem [{\citenamefont {Abbott}\ \emph
  {et~al.}(2018{\natexlab{b}})\citenamefont {Abbott} \emph
  {et~al.}}]{TheLIGOScientific:2017lwt}%
  \BibitemOpen
  \bibfield  {author} {\bibinfo {author} {\bibfnamefont {B.~P.}\ \bibnamefont
  {Abbott}} \emph {et~al.} (\bibinfo {collaboration} {LIGO Scientific,
  Virgo}),\ }\href {\doibase 10.1088/1361-6382/aaaafa} {\bibfield  {journal}
  {\bibinfo  {journal} {Class. Quant. Grav.}\ }\textbf {\bibinfo {volume}
  {35}},\ \bibinfo {pages} {065010} (\bibinfo {year} {2018}{\natexlab{b}})},\
  \Eprint {http://arxiv.org/abs/1710.02185} {arXiv:1710.02185 [gr-qc]}
  \BibitemShut {NoStop}%
\bibitem [{\citenamefont {Abbott}\ \emph
  {et~al.}(2018{\natexlab{c}})\citenamefont {Abbott} \emph
  {et~al.}}]{Abbott:2018oah}%
  \BibitemOpen
  \bibfield  {author} {\bibinfo {author} {\bibfnamefont {B.~P.}\ \bibnamefont
  {Abbott}} \emph {et~al.} (\bibinfo {collaboration} {LIGO Scientific,
  Virgo}),\ }\href {\doibase 10.1103/PhysRevLett.121.231103} {\bibfield
  {journal} {\bibinfo  {journal} {Phys. Rev. Lett.}\ }\textbf {\bibinfo
  {volume} {121}},\ \bibinfo {pages} {231103} (\bibinfo {year}
  {2018}{\natexlab{c}})},\ \Eprint {http://arxiv.org/abs/1808.04771}
  {arXiv:1808.04771 [astro-ph.CO]} \BibitemShut {NoStop}%
\bibitem [{\citenamefont {{Hall}}(2019{\natexlab{a}})}]{Hall_LIGOdoc}%
  \BibitemOpen
  \bibfield  {author} {\bibinfo {author} {\bibfnamefont {E.}~\bibnamefont
  {{Hall}}},\ }\href@noop {} {\enquote {\bibinfo {title} {{Horizon plots and
  noise curves for second- and third-generation detectors}},}\ }\bibinfo
  {howpublished} {LIGO Document T1800084-v5,
  \url{https://dcc.ligo.org/LIGO-T1800084-v5/public}} (\bibinfo {year}
  {2019}{\natexlab{a}})\BibitemShut {NoStop}%
\bibitem [{\citenamefont {{Hall}}(2019{\natexlab{b}})}]{Hall_code}%
  \BibitemOpen
  \bibfield  {author} {\bibinfo {author} {\bibfnamefont {E.}~\bibnamefont
  {{Hall}}},\ }\href@noop {} {\enquote {\bibinfo {title}
  {{{gw-horizon-plot}}},}\ }\bibinfo {howpublished}
  {\url{https://git.ligo.org/evan.hall/gw-horizon-plot}} (\bibinfo {year}
  {2019}{\natexlab{b}})\BibitemShut {NoStop}%
\bibitem [{\citenamefont {Fender}\ \emph {et~al.}(2013)\citenamefont {Fender},
  \citenamefont {Maccarone},\ and\ \citenamefont {Heywood}}]{Fender:2013ei}%
  \BibitemOpen
  \bibfield  {author} {\bibinfo {author} {\bibfnamefont {R.}~\bibnamefont
  {Fender}}, \bibinfo {author} {\bibfnamefont {T.}~\bibnamefont {Maccarone}}, \
  and\ \bibinfo {author} {\bibfnamefont {I.}~\bibnamefont {Heywood}},\ }\href
  {\doibase 10.1093/mnras/sts688} {\bibfield  {journal} {\bibinfo  {journal}
  {Mon. Not. Roy. Astron. Soc.}\ }\textbf {\bibinfo {volume} {430}},\ \bibinfo
  {pages} {1538} (\bibinfo {year} {2013})},\ \Eprint
  {http://arxiv.org/abs/1301.1341} {arXiv:1301.1341 [astro-ph.HE]} \BibitemShut
  {NoStop}%
\bibitem [{\citenamefont {Gaggero}\ \emph {et~al.}(2017)\citenamefont
  {Gaggero}, \citenamefont {Bertone}, \citenamefont {Calore}, \citenamefont
  {Connors}, \citenamefont {Lovell}, \citenamefont {Markoff},\ and\
  \citenamefont {Storm}}]{Gaggero:2016dpq}%
  \BibitemOpen
  \bibfield  {author} {\bibinfo {author} {\bibfnamefont {D.}~\bibnamefont
  {Gaggero}}, \bibinfo {author} {\bibfnamefont {G.}~\bibnamefont {Bertone}},
  \bibinfo {author} {\bibfnamefont {F.}~\bibnamefont {Calore}}, \bibinfo
  {author} {\bibfnamefont {R.~M.~T.}\ \bibnamefont {Connors}}, \bibinfo
  {author} {\bibfnamefont {M.}~\bibnamefont {Lovell}}, \bibinfo {author}
  {\bibfnamefont {S.}~\bibnamefont {Markoff}}, \ and\ \bibinfo {author}
  {\bibfnamefont {E.}~\bibnamefont {Storm}},\ }\href {\doibase
  10.1103/PhysRevLett.118.241101} {\bibfield  {journal} {\bibinfo  {journal}
  {Phys. Rev. Lett.}\ }\textbf {\bibinfo {volume} {118}},\ \bibinfo {pages}
  {241101} (\bibinfo {year} {2017})},\ \Eprint
  {http://arxiv.org/abs/1612.00457} {arXiv:1612.00457 [astro-ph.HE]}
  \BibitemShut {NoStop}%
\bibitem [{\citenamefont {Ivanov}\ \emph {et~al.}(2019)\citenamefont {Ivanov},
  \citenamefont {Lukash}, \citenamefont {Pilipenko},\ and\ \citenamefont
  {Pshirkov}}]{Ivanov:2019wkq}%
  \BibitemOpen
  \bibfield  {author} {\bibinfo {author} {\bibfnamefont {P.~B.}\ \bibnamefont
  {Ivanov}}, \bibinfo {author} {\bibfnamefont {V.~N.}\ \bibnamefont {Lukash}},
  \bibinfo {author} {\bibfnamefont {S.~V.}\ \bibnamefont {Pilipenko}}, \ and\
  \bibinfo {author} {\bibfnamefont {M.~S.}\ \bibnamefont {Pshirkov}},\ }\href
  {\doibase 10.1093/mnras/stz2206} {\bibfield  {journal} {\bibinfo  {journal}
  {Mon. Not. Roy. Astron. Soc.}\ }\textbf {\bibinfo {volume} {489}},\ \bibinfo
  {pages} {2038} (\bibinfo {year} {2019})},\ \Eprint
  {http://arxiv.org/abs/1905.04923} {arXiv:1905.04923 [astro-ph.HE]}
  \BibitemShut {NoStop}%
\bibitem [{\citenamefont {{Manshanden}}\ \emph {et~al.}(2018)\citenamefont
  {{Manshanden}}, \citenamefont {{Gaggero}}, \citenamefont {{Bertone}},
  \citenamefont {{Connors}},\ and\ \citenamefont
  {{Ricotti}}}]{2018arXiv181207967M}%
  \BibitemOpen
  \bibfield  {author} {\bibinfo {author} {\bibfnamefont {J.}~\bibnamefont
  {{Manshanden}}}, \bibinfo {author} {\bibfnamefont {D.}~\bibnamefont
  {{Gaggero}}}, \bibinfo {author} {\bibfnamefont {G.}~\bibnamefont
  {{Bertone}}}, \bibinfo {author} {\bibfnamefont {R.~M.~T.}\ \bibnamefont
  {{Connors}}}, \ and\ \bibinfo {author} {\bibfnamefont {M.}~\bibnamefont
  {{Ricotti}}},\ }\href@noop {} {\bibfield  {journal} {\bibinfo  {journal}
  {arXiv e-prints}\ } (\bibinfo {year} {2018})},\ \Eprint
  {http://arxiv.org/abs/1812.07967} {arXiv:1812.07967 [astro-ph.HE]}
  \BibitemShut {NoStop}%
\bibitem [{\citenamefont {Bull}\ \emph {et~al.}(2018)\citenamefont {Bull} \emph
  {et~al.}}]{Bull:2018lat}%
  \BibitemOpen
  \bibfield  {author} {\bibinfo {author} {\bibfnamefont {P.}~\bibnamefont
  {Bull}} \emph {et~al.},\ }\href@noop {} {\  (\bibinfo {year} {2018})},\
  \Eprint {http://arxiv.org/abs/1810.02680} {arXiv:1810.02680 [astro-ph.CO]}
  \BibitemShut {NoStop}%
\bibitem [{\citenamefont {Bertschinger}(1985)}]{Bertschinger:1985pd}%
  \BibitemOpen
  \bibfield  {author} {\bibinfo {author} {\bibfnamefont {E.}~\bibnamefont
  {Bertschinger}},\ }\href {\doibase 10.1086/191028} {\bibfield  {journal}
  {\bibinfo  {journal} {Astrophys. J. Suppl.}\ }\textbf {\bibinfo {volume}
  {58}},\ \bibinfo {pages} {39} (\bibinfo {year} {1985})}\BibitemShut {NoStop}%
\bibitem [{\citenamefont {Mack}\ \emph {et~al.}(2007)\citenamefont {Mack},
  \citenamefont {Ostriker},\ and\ \citenamefont {Ricotti}}]{Mack:2006gz}%
  \BibitemOpen
  \bibfield  {author} {\bibinfo {author} {\bibfnamefont {K.~J.}\ \bibnamefont
  {Mack}}, \bibinfo {author} {\bibfnamefont {J.~P.}\ \bibnamefont {Ostriker}},
  \ and\ \bibinfo {author} {\bibfnamefont {M.}~\bibnamefont {Ricotti}},\ }\href
  {\doibase 10.1086/518998} {\bibfield  {journal} {\bibinfo  {journal}
  {Astrophys. J.}\ }\textbf {\bibinfo {volume} {665}},\ \bibinfo {pages} {1277}
  (\bibinfo {year} {2007})},\ \Eprint {http://arxiv.org/abs/astro-ph/0608642}
  {arXiv:astro-ph/0608642 [astro-ph]} \BibitemShut {NoStop}%
\bibitem [{\citenamefont {Bertoni}\ \emph {et~al.}(2015)\citenamefont
  {Bertoni}, \citenamefont {Hooper},\ and\ \citenamefont
  {Linden}}]{Bertoni:2015mla}%
  \BibitemOpen
  \bibfield  {author} {\bibinfo {author} {\bibfnamefont {B.}~\bibnamefont
  {Bertoni}}, \bibinfo {author} {\bibfnamefont {D.}~\bibnamefont {Hooper}}, \
  and\ \bibinfo {author} {\bibfnamefont {T.}~\bibnamefont {Linden}},\ }\href
  {\doibase 10.1088/1475-7516/2015/12/035} {\bibfield  {journal} {\bibinfo
  {journal} {JCAP}\ }\textbf {\bibinfo {volume} {1512}},\ \bibinfo {pages}
  {035} (\bibinfo {year} {2015})},\ \Eprint {http://arxiv.org/abs/1504.02087}
  {arXiv:1504.02087 [astro-ph.HE]} \BibitemShut {NoStop}%
\bibitem [{\citenamefont {Hooper}\ and\ \citenamefont
  {Witte}(2017)}]{Hooper:2016cld}%
  \BibitemOpen
  \bibfield  {author} {\bibinfo {author} {\bibfnamefont {D.}~\bibnamefont
  {Hooper}}\ and\ \bibinfo {author} {\bibfnamefont {S.~J.}\ \bibnamefont
  {Witte}},\ }\href {\doibase 10.1088/1475-7516/2017/04/018} {\bibfield
  {journal} {\bibinfo  {journal} {JCAP}\ }\textbf {\bibinfo {volume} {1704}},\
  \bibinfo {pages} {018} (\bibinfo {year} {2017})},\ \Eprint
  {http://arxiv.org/abs/1610.07587} {arXiv:1610.07587 [astro-ph.HE]}
  \BibitemShut {NoStop}%
\bibitem [{\citenamefont {Schoonenberg}\ \emph {et~al.}(2016)\citenamefont
  {Schoonenberg}, \citenamefont {Gaskins}, \citenamefont {Bertone},\ and\
  \citenamefont {Diemand}}]{Schoonenberg:2016aml}%
  \BibitemOpen
  \bibfield  {author} {\bibinfo {author} {\bibfnamefont {D.}~\bibnamefont
  {Schoonenberg}}, \bibinfo {author} {\bibfnamefont {J.}~\bibnamefont
  {Gaskins}}, \bibinfo {author} {\bibfnamefont {G.}~\bibnamefont {Bertone}}, \
  and\ \bibinfo {author} {\bibfnamefont {J.}~\bibnamefont {Diemand}},\ }\href
  {\doibase 10.1088/1475-7516/2016/05/028} {\bibfield  {journal} {\bibinfo
  {journal} {JCAP}\ }\textbf {\bibinfo {volume} {1605}},\ \bibinfo {pages}
  {028} (\bibinfo {year} {2016})},\ \Eprint {http://arxiv.org/abs/1601.06781}
  {arXiv:1601.06781 [astro-ph.HE]} \BibitemShut {NoStop}%
\bibitem [{\citenamefont {Taylor}\ and\ \citenamefont
  {Silk}(2003)}]{Taylor:2002zd}%
  \BibitemOpen
  \bibfield  {author} {\bibinfo {author} {\bibfnamefont {J.~E.}\ \bibnamefont
  {Taylor}}\ and\ \bibinfo {author} {\bibfnamefont {J.}~\bibnamefont {Silk}},\
  }\href {\doibase 10.1046/j.1365-8711.2003.06201.x} {\bibfield  {journal}
  {\bibinfo  {journal} {Mon. Not. Roy. Astron. Soc.}\ }\textbf {\bibinfo
  {volume} {339}},\ \bibinfo {pages} {505} (\bibinfo {year} {2003})},\ \Eprint
  {http://arxiv.org/abs/astro-ph/0207299} {arXiv:astro-ph/0207299 [astro-ph]}
  \BibitemShut {NoStop}%
\bibitem [{\citenamefont {Ullio}\ \emph {et~al.}(2002)\citenamefont {Ullio},
  \citenamefont {Bergstrom}, \citenamefont {Edsjo},\ and\ \citenamefont
  {Lacey}}]{Ullio:2002pj}%
  \BibitemOpen
  \bibfield  {author} {\bibinfo {author} {\bibfnamefont {P.}~\bibnamefont
  {Ullio}}, \bibinfo {author} {\bibfnamefont {L.}~\bibnamefont {Bergstrom}},
  \bibinfo {author} {\bibfnamefont {J.}~\bibnamefont {Edsjo}}, \ and\ \bibinfo
  {author} {\bibfnamefont {C.~G.}\ \bibnamefont {Lacey}},\ }\href {\doibase
  10.1103/PhysRevD.66.123502} {\bibfield  {journal} {\bibinfo  {journal} {Phys.
  Rev.}\ }\textbf {\bibinfo {volume} {D66}},\ \bibinfo {pages} {123502}
  (\bibinfo {year} {2002})},\ \Eprint {http://arxiv.org/abs/astro-ph/0207125}
  {arXiv:astro-ph/0207125 [astro-ph]} \BibitemShut {NoStop}%
\bibitem [{\citenamefont {Ando}(2005)}]{Ando:2005hr}%
  \BibitemOpen
  \bibfield  {author} {\bibinfo {author} {\bibfnamefont {S.}~\bibnamefont
  {Ando}},\ }\href {\doibase 10.1103/PhysRevLett.94.171303} {\bibfield
  {journal} {\bibinfo  {journal} {Phys. Rev. Lett.}\ }\textbf {\bibinfo
  {volume} {94}},\ \bibinfo {pages} {171303} (\bibinfo {year} {2005})},\
  \Eprint {http://arxiv.org/abs/astro-ph/0503006} {arXiv:astro-ph/0503006
  [astro-ph]} \BibitemShut {NoStop}%
\bibitem [{\citenamefont {Ackermann}\ \emph
  {et~al.}(2015{\natexlab{a}})\citenamefont {Ackermann} \emph
  {et~al.}}]{Ackermann:2014usa}%
  \BibitemOpen
  \bibfield  {author} {\bibinfo {author} {\bibfnamefont {M.}~\bibnamefont
  {Ackermann}} \emph {et~al.} (\bibinfo {collaboration} {Fermi-LAT}),\ }\href
  {\doibase 10.1088/0004-637X/799/1/86} {\bibfield  {journal} {\bibinfo
  {journal} {Astrophys. J.}\ }\textbf {\bibinfo {volume} {799}},\ \bibinfo
  {pages} {86} (\bibinfo {year} {2015}{\natexlab{a}})},\ \Eprint
  {http://arxiv.org/abs/1410.3696} {arXiv:1410.3696 [astro-ph.HE]} \BibitemShut
  {NoStop}%
\bibitem [{\citenamefont {Athron}\ \emph
  {et~al.}(2017{\natexlab{a}})\citenamefont {Athron} \emph
  {et~al.}}]{Athron:2017yua}%
  \BibitemOpen
  \bibfield  {author} {\bibinfo {author} {\bibfnamefont {P.}~\bibnamefont
  {Athron}} \emph {et~al.} (\bibinfo {collaboration} {GAMBIT}),\ }\href
  {\doibase 10.1140/epjc/s10052-017-5196-8} {\bibfield  {journal} {\bibinfo
  {journal} {Eur. Phys. J.}\ }\textbf {\bibinfo {volume} {C77}},\ \bibinfo
  {pages} {879} (\bibinfo {year} {2017}{\natexlab{a}})},\ \Eprint
  {http://arxiv.org/abs/1705.07917} {arXiv:1705.07917 [hep-ph]} \BibitemShut
  {NoStop}%
\bibitem [{\citenamefont {Athron}\ \emph
  {et~al.}(2017{\natexlab{b}})\citenamefont {Athron} \emph
  {et~al.}}]{Athron:2017qdc}%
  \BibitemOpen
  \bibfield  {author} {\bibinfo {author} {\bibfnamefont {P.}~\bibnamefont
  {Athron}} \emph {et~al.} (\bibinfo {collaboration} {GAMBIT}),\ }\href
  {\doibase 10.1140/epjc/s10052-017-5167-0} {\bibfield  {journal} {\bibinfo
  {journal} {Eur. Phys. J.}\ }\textbf {\bibinfo {volume} {C77}},\ \bibinfo
  {pages} {824} (\bibinfo {year} {2017}{\natexlab{b}})},\ \Eprint
  {http://arxiv.org/abs/1705.07935} {arXiv:1705.07935 [hep-ph]} \BibitemShut
  {NoStop}%
\bibitem [{\citenamefont {Athron}\ \emph
  {et~al.}(2017{\natexlab{c}})\citenamefont {Athron} \emph
  {et~al.}}]{Athron:2017kgt}%
  \BibitemOpen
  \bibfield  {author} {\bibinfo {author} {\bibfnamefont {P.}~\bibnamefont
  {Athron}} \emph {et~al.} (\bibinfo {collaboration} {GAMBIT}),\ }\href
  {\doibase 10.1140/epjc/s10052-017-5113-1} {\bibfield  {journal} {\bibinfo
  {journal} {Eur. Phys. J.}\ }\textbf {\bibinfo {volume} {C77}},\ \bibinfo
  {pages} {568} (\bibinfo {year} {2017}{\natexlab{c}})},\ \Eprint
  {http://arxiv.org/abs/1705.07931} {arXiv:1705.07931 [hep-ph]} \BibitemShut
  {NoStop}%
\bibitem [{\citenamefont {Griest}\ and\ \citenamefont
  {Kamionkowski}(1990)}]{UnitarityGriestKamionkowski}%
  \BibitemOpen
  \bibfield  {author} {\bibinfo {author} {\bibfnamefont {K.}~\bibnamefont
  {Griest}}\ and\ \bibinfo {author} {\bibfnamefont {M.}~\bibnamefont
  {Kamionkowski}},\ }\href {\doibase 10.1103/PhysRevLett.64.615} {\bibfield
  {journal} {\bibinfo  {journal} {Phys. Rev. Lett.}\ }\textbf {\bibinfo
  {volume} {64}},\ \bibinfo {pages} {615} (\bibinfo {year} {1990})}\BibitemShut
  {NoStop}%
\bibitem [{\citenamefont {Gómez-Vargas}\ \emph {et~al.}(2017)\citenamefont
  {Gómez-Vargas}, \citenamefont {López-Fogliani}, \citenamefont {Muñoz},
  \citenamefont {Perez},\ and\ \citenamefont {Ruiz~de
  Austri}}]{Gomez-Vargas:2016ocf}%
  \BibitemOpen
  \bibfield  {author} {\bibinfo {author} {\bibfnamefont {G.~A.}\ \bibnamefont
  {Gómez-Vargas}}, \bibinfo {author} {\bibfnamefont {D.~E.}\ \bibnamefont
  {López-Fogliani}}, \bibinfo {author} {\bibfnamefont {C.}~\bibnamefont
  {Muñoz}}, \bibinfo {author} {\bibfnamefont {A.~D.}\ \bibnamefont {Perez}}, \
  and\ \bibinfo {author} {\bibfnamefont {R.}~\bibnamefont {Ruiz~de Austri}},\
  }\href {\doibase 10.1088/1475-7516/2017/03/047} {\bibfield  {journal}
  {\bibinfo  {journal} {JCAP}\ }\textbf {\bibinfo {volume} {1703}},\ \bibinfo
  {pages} {047} (\bibinfo {year} {2017})},\ \Eprint
  {http://arxiv.org/abs/1608.08640} {arXiv:1608.08640 [hep-ph]} \BibitemShut
  {NoStop}%
\bibitem [{\citenamefont {Steffen}(2006)}]{Steffen:2006hw}%
  \BibitemOpen
  \bibfield  {author} {\bibinfo {author} {\bibfnamefont {F.~D.}\ \bibnamefont
  {Steffen}},\ }\href {\doibase 10.1088/1475-7516/2006/09/001} {\bibfield
  {journal} {\bibinfo  {journal} {JCAP}\ }\textbf {\bibinfo {volume} {0609}},\
  \bibinfo {pages} {001} (\bibinfo {year} {2006})},\ \Eprint
  {http://arxiv.org/abs/hep-ph/0605306} {arXiv:hep-ph/0605306 [hep-ph]}
  \BibitemShut {NoStop}%
\bibitem [{\citenamefont {Petraki}\ and\ \citenamefont
  {Volkas}(2013)}]{Petraki:2013wwa}%
  \BibitemOpen
  \bibfield  {author} {\bibinfo {author} {\bibfnamefont {K.}~\bibnamefont
  {Petraki}}\ and\ \bibinfo {author} {\bibfnamefont {R.~R.}\ \bibnamefont
  {Volkas}},\ }\href {\doibase 10.1142/S0217751X13300287} {\bibfield  {journal}
  {\bibinfo  {journal} {Int. J. Mod. Phys.}\ }\textbf {\bibinfo {volume}
  {A28}},\ \bibinfo {pages} {1330028} (\bibinfo {year} {2013})},\ \Eprint
  {http://arxiv.org/abs/1305.4939} {arXiv:1305.4939 [hep-ph]} \BibitemShut
  {NoStop}%
\bibitem [{\citenamefont {Zurek}(2014)}]{Zurek:2013wia}%
  \BibitemOpen
  \bibfield  {author} {\bibinfo {author} {\bibfnamefont {K.~M.}\ \bibnamefont
  {Zurek}},\ }\href {\doibase 10.1016/j.physrep.2013.12.001} {\bibfield
  {journal} {\bibinfo  {journal} {Phys. Rept.}\ }\textbf {\bibinfo {volume}
  {537}},\ \bibinfo {pages} {91} (\bibinfo {year} {2014})},\ \Eprint
  {http://arxiv.org/abs/1308.0338} {arXiv:1308.0338 [hep-ph]} \BibitemShut
  {NoStop}%
\bibitem [{\citenamefont {Abramowski}\ \emph {et~al.}(2011)\citenamefont
  {Abramowski} \emph {et~al.}}]{Abramowski:2011hc}%
  \BibitemOpen
  \bibfield  {author} {\bibinfo {author} {\bibfnamefont {A.}~\bibnamefont
  {Abramowski}} \emph {et~al.} (\bibinfo {collaboration} {H.E.S.S.}),\ }\href
  {\doibase 10.1103/PhysRevLett.106.161301} {\bibfield  {journal} {\bibinfo
  {journal} {Phys. Rev. Lett.}\ }\textbf {\bibinfo {volume} {106}},\ \bibinfo
  {pages} {161301} (\bibinfo {year} {2011})},\ \Eprint
  {http://arxiv.org/abs/1103.3266} {arXiv:1103.3266 [astro-ph.HE]} \BibitemShut
  {NoStop}%
\bibitem [{\citenamefont {{Ackermann}}\ \emph {et~al.}(2012)\citenamefont
  {{Ackermann}} \emph {et~al.}}]{2012ApJ...761...91A}%
  \BibitemOpen
  \bibfield  {author} {\bibinfo {author} {\bibfnamefont {M.}~\bibnamefont
  {{Ackermann}}} \emph {et~al.},\ }\href {\doibase 10.1088/0004-637X/761/2/91}
  {\bibfield  {journal} {\bibinfo  {journal} {Astrophys. J}\ }\textbf {\bibinfo
  {volume} {761}},\ \bibinfo {eid} {91} (\bibinfo {year} {2012})},\ \Eprint
  {http://arxiv.org/abs/1205.6474} {arXiv:1205.6474 [astro-ph.CO]} \BibitemShut
  {NoStop}%
\bibitem [{\citenamefont {Ackermann}\ \emph
  {et~al.}(2015{\natexlab{b}})\citenamefont {Ackermann} \emph
  {et~al.}}]{Ackermann:2015zua}%
  \BibitemOpen
  \bibfield  {author} {\bibinfo {author} {\bibfnamefont {M.}~\bibnamefont
  {Ackermann}} \emph {et~al.} (\bibinfo {collaboration} {Fermi-LAT}),\ }\href
  {\doibase 10.1103/PhysRevLett.115.231301} {\bibfield  {journal} {\bibinfo
  {journal} {Phys. Rev. Lett.}\ }\textbf {\bibinfo {volume} {115}},\ \bibinfo
  {pages} {231301} (\bibinfo {year} {2015}{\natexlab{b}})},\ \Eprint
  {http://arxiv.org/abs/1503.02641} {arXiv:1503.02641 [astro-ph.HE]}
  \BibitemShut {NoStop}%
\bibitem [{\citenamefont {Abdallah}\ \emph {et~al.}(2018)\citenamefont
  {Abdallah} \emph {et~al.}}]{Abdallah:2018qtu}%
  \BibitemOpen
  \bibfield  {author} {\bibinfo {author} {\bibfnamefont {H.}~\bibnamefont
  {Abdallah}} \emph {et~al.} (\bibinfo {collaboration} {HESS}),\ }\href
  {\doibase 10.1103/PhysRevLett.120.201101} {\bibfield  {journal} {\bibinfo
  {journal} {Phys. Rev. Lett.}\ }\textbf {\bibinfo {volume} {120}},\ \bibinfo
  {pages} {201101} (\bibinfo {year} {2018})},\ \Eprint
  {http://arxiv.org/abs/1805.05741} {arXiv:1805.05741 [astro-ph.HE]}
  \BibitemShut {NoStop}%
\bibitem [{\citenamefont {Rohatgi}(2018)}]{WebPlotDigitizer}%
  \BibitemOpen
  \bibfield  {author} {\bibinfo {author} {\bibfnamefont {A.}~\bibnamefont
  {Rohatgi}},\ }\href@noop {} {\enquote {\bibinfo {title} {Webplotdigitizer},}\
  }\bibinfo {howpublished} {https://automeris.io/WebPlotDigitizer} (\bibinfo
  {year} {2018})\BibitemShut {NoStop}%
\bibitem [{\citenamefont {Cirelli}\ \emph {et~al.}(2011)\citenamefont
  {Cirelli}, \citenamefont {Corcella}, \citenamefont {Hektor}, \citenamefont
  {Hutsi}, \citenamefont {Kadastik}, \citenamefont {Panci}, \citenamefont
  {Raidal}, \citenamefont {Sala},\ and\ \citenamefont
  {Strumia}}]{Cirelli:2010xx}%
  \BibitemOpen
  \bibfield  {author} {\bibinfo {author} {\bibfnamefont {M.}~\bibnamefont
  {Cirelli}}, \bibinfo {author} {\bibfnamefont {G.}~\bibnamefont {Corcella}},
  \bibinfo {author} {\bibfnamefont {A.}~\bibnamefont {Hektor}}, \bibinfo
  {author} {\bibfnamefont {G.}~\bibnamefont {Hutsi}}, \bibinfo {author}
  {\bibfnamefont {M.}~\bibnamefont {Kadastik}}, \bibinfo {author}
  {\bibfnamefont {P.}~\bibnamefont {Panci}}, \bibinfo {author} {\bibfnamefont
  {M.}~\bibnamefont {Raidal}}, \bibinfo {author} {\bibfnamefont
  {F.}~\bibnamefont {Sala}}, \ and\ \bibinfo {author} {\bibfnamefont
  {A.}~\bibnamefont {Strumia}},\ }\href {\doibase
  10.1088/1475-7516/2012/10/E01, 10.1088/1475-7516/2011/03/051} {\bibfield
  {journal} {\bibinfo  {journal} {JCAP}\ }\textbf {\bibinfo {volume} {1103}},\
  \bibinfo {pages} {051} (\bibinfo {year} {2011})},\ \bibinfo {note} {[Erratum:
  JCAP1210,E01(2012)]},\ \Eprint {http://arxiv.org/abs/1012.4515}
  {arXiv:1012.4515 [hep-ph]} \BibitemShut {NoStop}%
\bibitem [{\citenamefont {Ando}\ and\ \citenamefont
  {Ishiwata}(2015)}]{Ando:2015qda}%
  \BibitemOpen
  \bibfield  {author} {\bibinfo {author} {\bibfnamefont {S.}~\bibnamefont
  {Ando}}\ and\ \bibinfo {author} {\bibfnamefont {K.}~\bibnamefont
  {Ishiwata}},\ }\href {\doibase 10.1088/1475-7516/2015/05/024} {\bibfield
  {journal} {\bibinfo  {journal} {JCAP}\ }\textbf {\bibinfo {volume} {1505}},\
  \bibinfo {pages} {024} (\bibinfo {year} {2015})},\ \Eprint
  {http://arxiv.org/abs/1502.02007} {arXiv:1502.02007 [astro-ph.CO]}
  \BibitemShut {NoStop}%
\end{thebibliography}%

\appendix

\section{Gamma ray constraints}
\label{app:gamma_ray_constraints}

Here we give the details of our procedure for constraining $\sv$ in each PBH detection scenario using their diffuse gamma-ray emission and by treating them as gamma-ray point sources. For each detection scenario, as point estimates for $f_\pbh$ we conservatively use the 5th percentile for $\hat{f}_\pbh$ computed using the posteriors $P(f_\pbh | N)$ found above. These values are collected in Table~\ref{tab:f_5th_percentile}.

The key ingredient for both the diffuse and point source analyses is the WIMP annihilation rate around a PBH. As described in the main text, the ultracompact minihalos (UCMHs) surrounding PBHs have $\rho \propto r^{-9/4}$ DM halos. Since the maximum-possible WIMP density at present is $\rho_\u{max} = m_\dm / \sv \, t_0$ (where $t_0$ is the age of the universe),\footnote{This is conservative since $\rho_\u{max}$ was not derived for particles moving along radial trajectories~\cite{Lacki:2010zf}.} the UCMH density profile is constant within the cutoff radius~\cite{Adamek:2019gns}
\begin{align}
    r_\u{cut} &= 9.1\ee{-8}\us{kpc} \left( \frac{\sv}{3\ee{-26}\us{cm}^3/\u{s}} \right)^{4/9}\\
    &\hspace{1.5cm} \times \left( \frac{m_\dm}{100\us{GeV}} \right)^{-4/9} \left( \frac{M_\pbh}{M_\odot} \right)^{1/3}\,.
\end{align}
The WIMP annihilation rate in a UCMH is obtained by integrating the UCMH density squared over all space~\footnote{Changing the bound of the radial integral to the minihalo's tidal radius (see~\cite{Adamek:2019gns}, Sec.~2-E) does not appreciably change the annihilation rate.}
\begin{align}
    \Gamma &= \frac{4 \pi \sv \rho_\u{max}^2 r_\u{cut}^3}{2 m_\dm}\,.
\end{align}

We note that the maximum density (and the corresponding cutoff) apply only to the WIMP component of Dark Matter in the UCMH; we therefore now consider the case of a sub-dominant WIMP. In standard indirect detection analyses the gamma-ray flux depends separately on the integral of the WIMP density squared in the observing region (the $J$-factor) and particle physics factors, giving an overall dependence on $f_\dm^2 \sv$ for under-abundant species making up a fraction $f_\dm$ of the DM. In the UCMH scenario, the dependence is more complicated since the WIMP density profile depends on the particle physics through $\sv$. Since $r_\u{cut}$ is the radius at which the WIMP density attains its maximum value and $\rho_\dm(r) = f_\dm \rho_\cdm(r)$, we have $r_\u{cut} \propto f_\dm^{4/9}$. In contrast, $\rho_\u{max}$ does not depend on $f_\dm$. This means that the WIMP annihilation rate for an under-abundant species scales as
\begin{align}
    \Gamma \propto \left[ f_\chi^4 \sv \right]^{1/3}.
\end{align}

Finally, we note that we aim to constrain the \textit{zero-velocity limit} of the WIMP annihilation cross section $\sv$, which is typically the most relevant for annihilation signals today. We compare in Fig.~\ref{fig:sv_limits} with the corresponding quantity calculated for different theories of Weak-scale New Physics. This comparison (along with our conclusions about the constraining power of a future PBH detection) therefore does not depend strongly on assumptions about the velocity dependence of the cross section.

{\renewcommand{\arraystretch}{1.2}
\begin{table}[tbh!]
    \centering
    \begin{tabular}{c | c | c c c}
        \toprule
                                     & Prior    & $0.5~M_\odot$ & $10~M_\odot$  & $100~M_\odot$\\
        \colrule
        \multirow{2}{*}{$N_\u{min}$} & Log-flat & $1.49\ee{-3}$ & $7.28\ee{-6}$ & $2.56\ee{-5}$\\
                                     & Jeffreys & $2.81\ee{-3}$ & $1.32\ee{-5}$ & $2.82\ee{-5}$\\
        \colrule
        \multirow{2}{*}{$N_\u{max}$} & Log-flat & $1.10\ee{-1}$ & $4.17\ee{-3}$ & $3.34\ee{-4}$\\
                                     & Jeffreys & $1.10\ee{-1}$ & $4.17\ee{-3}$ & $3.36\ee{-4}$\\
        \botrule
    \end{tabular}
    \caption{\textbf{Point estimates for $f_\pbh$.} The entries in this table are the 5th percentiles of the posteriors $P(f_\pbh|N)$ for each scenario in Table~\ref{tab:Detections}. The second column indicates the prior for the merger rate (in the case of gravitational wave scenarios) or mean number of expected observations (for the SKA scenario).}
    \label{tab:f_5th_percentile}
\end{table}}

\subsection{PBHs as gamma-ray point sources}
\label{sub:point_source_constraints}

The natural test statistic for this analysis is $N_\gamma$, the number of PBHs that pass the Fermi analysis cuts and appear in the 3FGL unassociated point source list. Since there are 19 observed unassociated point sources, to set an upper bound on $\sv$ at the $\alpha=0.95$ level we require the probability of observing at least 19 unassociated point sources to be equal to the corresponding $p$-value of 0.05. This probability is the cumulative distribution function for a binomial distribution:
\begin{align}
    \operatorname{Pr}(N_\gamma \geq 19) = \sum_{N_\gamma = 19}^{N_\u{MW}} B(N_\gamma | N_\u{MW}, p_\gamma) = 0.05,
\end{align}
 where $N_\u{MW} \equiv \lfloor f_\pbh (3\ee{12}~M_\odot) / M_\pbh \rfloor$ is the number of PBHs in the Milky Way and $p_\gamma(\sv, \hat{f}_\pbh, M_\pbh, m_\dm)$ is the probability that a given PBH appears in the unassociated point source catalogue. With all other parameters fixed, this defines the upper bound for $\sv$, which must be determined numerically. We neglect the possibility that astrophysical sources appear in the unassociated point source catalogue since that would only strengthen our bounds.
 
 We compute $p_\gamma$ using a Monte Carlo procedure. First, a PBH's position is randomly sampled. The PBH spatial distribution is assumed to track the Galactic DM distribution, here taken to be the Einasto profile
\begin{align}
    \rho_\u{Ein}(r) &= \rho_s \exp \left\{ - \frac{2}{\alpha} \left[ \left( \frac{r}{r_s} \right)^\alpha - 1 \right] \right\},
\end{align}
where $r$ is the PBH's galactocentric distance and the halo parameters are $(\rho_s, r_s, \alpha) = (0.033\us{GeV/cm}^3,~28.44\us{kpc},~0.17)$~\cite{Cirelli:2010xx}. This parameterization is convenient since the quantity $r^\alpha$ can be efficiently sampled because its PDF is a gamma distribution with scale $\theta_r = \alpha r_s^\alpha / 2$ and shape $k_r = 3 / \alpha$; the PBH's Galactic longitude and sine of its latitude are sampled uniformly.

Next, we assume that any PBH with sufficiently large integrated flux above 1 GeV lying far enough outside the Galactic plane appears in the unassociated point source catalogue~\cite{Hooper:2016cld}:
\begin{align}
    \label{eq:pt_src_cuts}
    |b| > 20^\circ, \ \Phi_{> 1\us{GeV}} > 7\ee{-10}\us{cm}^{-2}\us{s}^{-1}.
\end{align}
Using the annihilation rate defined above, the differential gamma-ray flux from the PBH is
\begin{align}
    \label{eq:pbh_ps_flux}
    \phi_\pbh^\u{point}(E) &= \frac{\Gamma}{4 \pi d^2} \frac{\mathrm{d}N_\gamma}{\mathrm{d}E}(E)
\end{align}
where $\mathrm{d}N/\mathrm{d}E$ is the energy spectrum of photons per DM annihilation, which is easily integrated over energy. The final MC estimate for $p_\gamma$ is obtained by repeating this procedure and multiplying the fraction of sampled PBHs passing the detectability cuts by $N_\u{MW}$.

Since there are roughly $N_\u{MW} = 1\ee{5} - 2\ee{11}$ PBHs in the Milky Way (depending on $N_\pbh$ and $M_\pbh$), $\sv$ and thus $p_\gamma$ must be very small to give fewer than $N_{\mathcal{U}} = 19$ point sources. A naive Monte Carlo simulation is extremely inefficient for such small cross sections, but with importance sampling we can obtain more accurate results for lower computational cost. In detail, Eq.~\ref{eq:pbh_ps_flux} and the flux threshold from Eq.~\ref{eq:pt_src_cuts} define the maximum distance $d_\u{max}$ at which a PBH is detectable, which means we should only sample PBHs within this distance of Earth. Denoting the distance from the Galactic Center to Earth as $d_\oplus$, we sample the PBH galactic longitude uniformly from $[-\arcsin\frac{d_\u{max}}{d_\oplus}, \arcsin\frac{d_\u{max}}{d_\oplus}]$ and the sine of its galactic latitude uniformly from $[-d_\u{max} / d_\oplus, d_\u{max} / d_\oplus]$. Furthermore, inverse CDF sampling allows us to sample $r$ from the interval $[d_\oplus - d_\u{max}, d_\oplus + d_\u{max}]$. The Monte Carlo samples must be reweighted to account for the restricted sampling volume.

\subsection{Diffuse gamma rays from PBHs}
\label{sub:diffuse_constraints}

The photons emitted from DM halos around PBHs located across all redshifts contribute to the extragalactic gamma-ray background (EGB). Taking the gamma-ray flux in each of the 26 energy bins Fermi uses for their EGB analysis to be normally distributed, the sum of the squared, standardized fluxes follows a $\chi^2$ distribution, motivating the use of this quantity as the test statistic:
\begin{align}
    \chi^2 = \sum_{i=1}^{26} \left( \frac{\phi_\u{obs}^\u{ex}(E_i) - [\phi_\u{PBH}^\u{ex}(E_i) + \phi_\u{bg}^\u{ex}(E_i)]}{\sigma_i} \right)^2.
\end{align}
The $i$th bin energies and error bars are denoted by $E_i$ and $\sigma_i$ and the differential fluxes in the numerator are the observed flux, flux from UCMHs surrounding PBHs and background flux respectively.

The PBH flux depends on the PBH and Dark Matter parameters and is well-known:~\cite{Boucenna:2017ghj,Ando:2015qda,Cirelli:2010xx}
\begin{align}
    \label{eq:pbh_ex_flux}
    \phi_\pbh^\u{ex}(E) &= \frac{\Gamma}{4 \pi} \frac{\hat{f}_\pbh \, \rho_\cdm}{M_\pbh} \int_0^z \mathrm{d}z\ \frac{e^{-\tau(E, z)}}{H(z)} \frac{dN_\gamma}{dE}\left( (1+z) E \right).
\end{align}
The optical depth $\tau(E,z)$ accounts for attenuation of gamma rays emitted from redshift $z$ and observed with energy $E$ due primarily to pair production on baryonic matter and scattering off the extragalactic background light, for which we use the tables from~\cite{Cirelli:2010xx}. The cosmologically averaged DM density is $\rho_\cdm \approx 30~M_\odot / \u{kpc}^3$.

Instead of modeling the astrophysical contributions to the EGB from e.g.\ blazars, star-forming galaxies and misaligned active galactic nuclei, we set $\phi_\u{bg}^\u{ex}(E_i) = \phi_\u{obs}^\u{ex}(E_i)$. This yields less stringent limits than Ref.~\cite{Ando:2015qda} and subsequent mixed DM-PBH analyses utilizing their results~\cite{Boucenna:2017ghj,Adamek:2019gns} since their fiducial EGB background model is in tension with the Fermi observations. With this simplification the \emph{observed} test statistic reduces to $\chi_\u{obs}^2 = \sum_{i=1}^{26} \left[ \phi_\u{PBH}^\u{ex}(E_i) / \sigma_i \right]^2$.

To set a limit at the level $\alpha=0.95$ requires setting the probability of the test statistic exceeding the observed value to the corresponding $p$-value of 0.05:
\begin{align}
    \operatorname{Pr}(\chi^2 \geq \chi_\u{obs}^2) = \int_{\chi_\u{obs}^2}^\infty  P(\chi^2) \,\mathrm{d}\chi^2 &= 0.05.
\end{align}
This implies $\chi_\u{obs}^2 = 38.9$, the relevant critical value for 26 degrees of freedom. Substituting Eq.~\ref{eq:pbh_ex_flux} into the definition of $\chi^2_\u{obs}$ yields the bound on $\sv$. \\~\\

\begin{figure*}[p!]
    \centering
    \includegraphics[width=\textwidth]{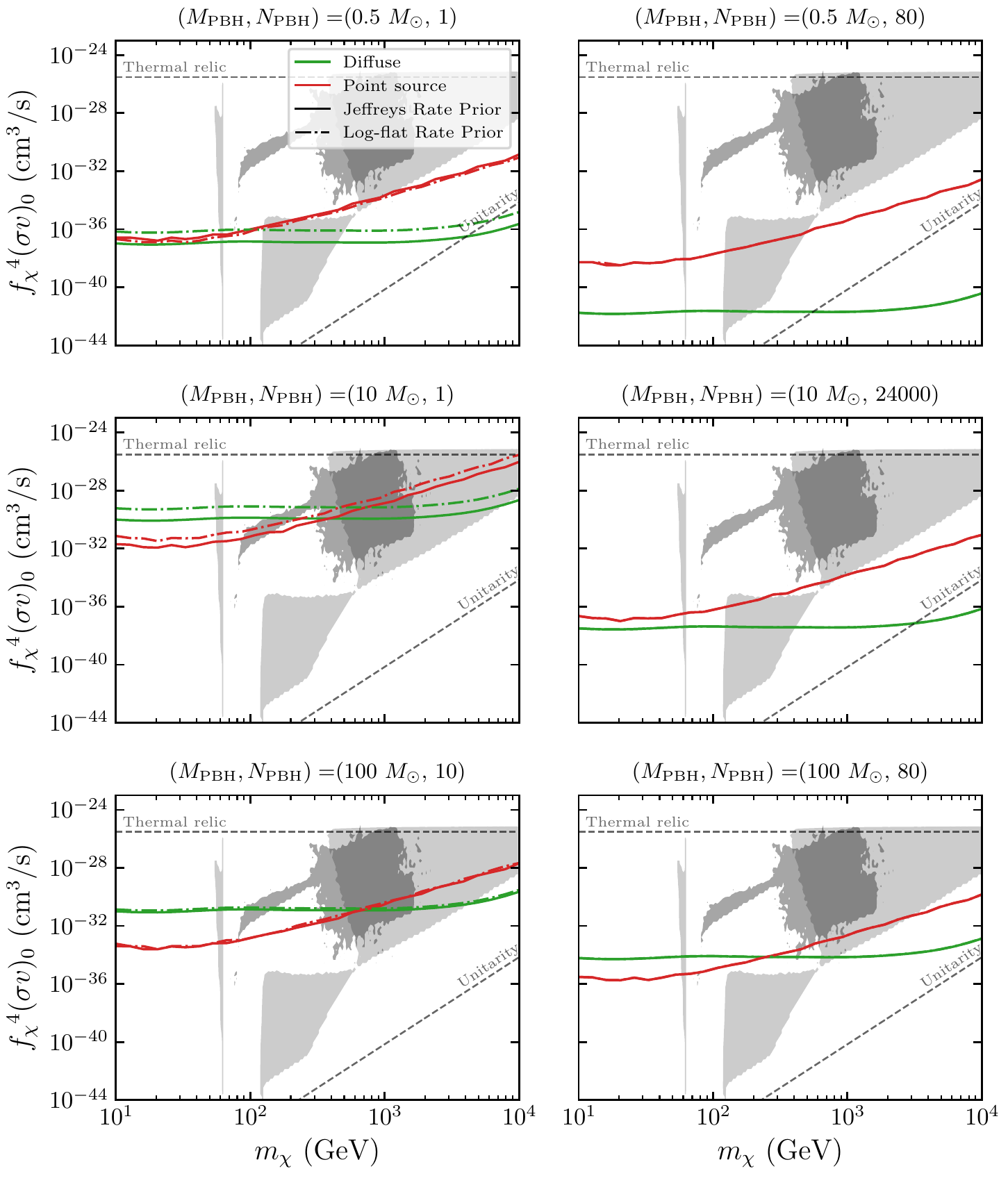}
    \caption{\textbf{Diffuse and point-source 95\% CL upper limits on $\sv$.} Each row corresponds to a different PBH detection scenario: LIGO O3 (top), Einstein Telescope (middle) and SKA (bottom). Green and red lines show constraints derived from diffuse extragalactic emission and point sources respectively. Solid lines show constraints assuming a Jeffreys prior on the rate of PBH observations, while dot-dashed lines show results for the more conservative log-flat prior. Note that in the right panels, the solid and dot-dashed lines are largely indistinguishable. In Fig.~\ref{fig:sv_limits}, we show results for the Jeffreys prior, taking the stronger of the diffuse and point source constraints at a given DM mass. \href{https://github.com/adam-coogan/pbhs_vs_wimps/blob/master/plot_frequentist_limits.py}{\faFileCodeO}}
    \label{fig:sv_limits_ps_diff_comparison}
\end{figure*}

In Fig.~\ref{fig:sv_limits_ps_diff_comparison}, we show separately the diffuse and point source constraints arising in each of the detection scenarios we have considered. 

The point source constraints are most important for small $N_\pbh$ while the diffuse constraints dominate for the values of $N_\u{max}$ in Table~\ref{tab:Detections}. The difference in scaling is because the number of PBHs passing the integrated flux cut is proportional to $f_\mathrm{PBH}$ times $p_\gamma$ (the probability of a PBH lying in the tail of the integrated flux distribution), which is a very sensitive function of $\Gamma \propto \sv^{1/3}$ since the distribution is roughly log-normal. This means that holding $N_\gamma$ fixed while increasing $f_\mathrm{PBH}$ translates into a small decrease in $\sv$. In contrast, the diffuse extragalactic flux from PBHs scales less strongly with $\sv$ as $\phi_\pbh^\u{ex} \propto f_\pbh \sv^{1/3}$, so the cross section constraint scales as $f_\pbh^3$. These different scalings for the diffuse and point-source cross section bounds are exhibited in Fig.~\ref{fig:sv_limits_ps_diff_comparison}.

\subsection{Dependence of results on priors}
\label{app:prior_dependence}

In Fig.~\ref{fig:sv_limits_ps_diff_comparison}, we also show the projected constraints on $\sv$ for two different choices of prior on the mean PBH detection rate. Recall that to obtain the posterior for $f_\pbh$, we must choose a prior for the mean number of PBH candidates $\Lambda$ which we expect to observe. Solid lines in Fig.~\ref{fig:sv_limits_ps_diff_comparison} show the projected bounds assuming a Jeffreys' prior on the PBH rate. For a Poisson process, the Jeffreys' prior (which is invariant under reparametrization) is given by $\mathrm{Pr}(\Lambda)\propto 1/\Lambda^{1/2}$. Dot-dashed lines show instead the results assuming a log-flat prior on the mean PBH detection rate, $\mathrm{Pr}(\Lambda)\propto 1/\Lambda$. The Jeffreys' prior is that assumed by LIGO/Virgo \cite{Abbott:2016nhf,Abbott:2016drs} and corresponds to the results we show in Fig.~\ref{fig:sv_limits}. Instead, the log-flat prior pushes the posterior towards a smaller detection rate, smaller values of $f_\pbh$ and therefore weaker constraints on $\sv$.

We see that with only a single PBH detection in gravitational waves (upper-left and middle-left panels), the log-flat prior may weaken the possible constraints by up to an order of magnitude. This is because with a small number of detections, the posterior on $f_\pbh$ is still rather broad (see Fig.~\ref{fig:Constraints_f}) and so a change to a more conservative prior can noticeably lift the constraints. For a larger number of observations (right panels of Fig.~\ref{fig:sv_limits_ps_diff_comparison}), the results are insensitive to the choice of prior (the solid and dot-dashed lines are almost indistinguishable). This is to be expected; in this case, the posterior is dominated by the Poissonian likelihood for observing $N_\mathrm{max}$ PBH candidates.

Even in the scenarios where a change of prior on the rate weakens the projected constraints, our main conclusions are not substantively changed: even a small number of PBH detections can place stringent constraints on thermal WIMP Dark Matter and, more generally, theories of New Physics at the Weak Scale.

\end{document}